\pdfoutput=1
\documentclass[12pt,a4paper]{article}

\usepackage{jheppub}

\usepackage{graphicx}	% Including figure files
\usepackage{amssymb} % for \geqslant and \leqslant
\usepackage{amsmath} % for \text{} in equations and 'align' environment

\usepackage{har2nat} % "natbib" is loaded automatically
\usepackage[]{natbib}

\usepackage{soulutf8}

\title{Theory of magnetic flux tubes in strong fields and the phenomenon of dark matter axions identical to solar axions}

\author[a]{Vitaliy D. Rusov\note{Corresponding author: Vitaliy D. Rusov, siiis@te.net.ua}}
\author[a]{Vladimir P. Smolyar}
\author[b]{Margarita E. Beglaryan}

\affiliation[a]{Department of Theoretical and Experimental Nuclear Physics,\\Odessa National Polytechnic University, Odessa, Ukraine}
\affiliation[b]{Department of Computer Technology and Applied Mathematics, Kuban State University,\\Krasnodar, Russia}

\emailAdd{siiis@te.net.ua}

\abstract{We develop the general laws of the theory of the almost empty anchored magnetic flux tubes (MFT) with $B\sim 10^7~G$, starting from tachocline to the surface of the Sun. The main result of this theory is the formation of the solar axions and the magnetic O-loop inside the MFT near the tachocline. In this magnetic O-loop (based on the Kolmogorov turbulent cascade) the axions are converted to photons, producing the axion origin photons near the bottom of the convective zone, i.e. near the tachocline. On the other hand, high-energy photons from the radiation zone through the axion-photon oscillations in the O-loop inside the MFT near the tachocline produce the so-called axions of photon origin under the sunspot. This means that at such strong magnetic fields the Parker-Biermann cooling effect of MFT develops due to the ``disappearance'' of the Parker's convective heat transport, and consequently, the temperature in the lower part of the magnetic tube with the help of the axions of photonic origin from the photon-axion oscillations in the O-loop near the tachocline.
As a result, a free path for photons of axion origin opens from the tachocline to the photosphere!

We show that the width of the ring between the magnetic wall of the flux tube and the magnetic O-loop near the tachocline, is the distance between the Parker-Biermann cooling effect, which forms the flux tube buoyancy, and the convective heating source $(dQ/dt)_2$, which, despite the neutral buoyancy of the magnetic tube ring, repeats the upstream tube identically. In other words, the source of convective heating $(dQ/dt)_2$ appears on the surface of the Sun as a result of the initiation (due to the Parker-Biermann cooling effect) of adiabatic growth of an undulatory instability in toroidal flux tubes which initially were in mechanical equilibrium (neutral buoyancy) in the stable overshoot layer.

\newpage

A very beautiful problem was solved this way. The high-energy photons passing from the radiation zone through the horizontal field of the O-loop near the tachocline turn into axions (see inverse oscillations $\gamma + \vec{B} \rightarrow \vec{B} + a$), which almost completely eliminates the radiation heating $(dQ/dt)_1$ of the almost empty magnetic flux tube. A certain flux of photons coming from the radiation zone through the tachocline, passes through the ``ring'' of a strong magnetic tube by virtue of convective heating $(dQ/dt)_2$. It allows to determine both the velocity and the lifetime of the magnetic flux tube (with $B\sim 10^7~G$) before the reconnection, from tachocline to the surface of the Sun, as well as the rate of the magnetic flux tube reconnection (with $B\sim 10^5~G$) near the tachocline $V_{rec}$. The latter is determined only by the rise time of the magnetic loop and the disappearance of the spot from the surface of the Sun.

Finally, we can show that the formation of sunspot cycles is equivalent to the number of cycles from the MFT, which coincides with the observational data of the Joy's law, and that both effects are the manifestations of dark matter (DM) -- solar axions in the core of the Sun, whose modulations are controlled by the anticorrelated 11-year modulation of asymmetric dark matter (ADM) density in the solar interior.
}

\date{}

\begin{document}

\maketitle
%\flushbottom

%\begin{keywords}
%solar axions -- magnetic tubes -- sunspots -- thermomagnetic Eitingshausen-Nernst effect -- holographic principle of quantum gravity
%\end{keywords}

\newlength{\myIndent}
\setlength{\myIndent}{\parindent}

\section{Introduction and motivation}

It is known that the unsolved problem of energy transport by magnetic flux
tubes at the same time represents another unsolved problem related to the
sunspot darkness (see 2.2 in \cite{Rempel2011}). Of all the known concepts 
playing a noticeable role in understanding the connection between the energy
transport and sunspot darkness, let us consider the most significant theory,
in our view. It is based on the Parker-Biermann cooling
effect~\citep{Parker1955a,Biermann1941,Parker1979b} and originates from the
early works of \cite{Biermann1941} and \cite{Alfven1942}.

As you know, the Parker-Biermann cooling
effect~\citep{Parker1955a,Biermann1941,Parker1979b}, which plays a role in our
current understanding, originates from \cite{Biermann1941} and
\cite{Alfven1942}: in a highly ionized plasma, the electrical
conductivity can be so large that the magnetic fields are frozen into the
plasma. Biermann realized that the magnetic field in the spots themselves can
be the cause of their coolness -- it is colder because the magnetic field
suppresses the convective heat transport. Hence, the darkness of the spot is
due to a decrease in surface brightness.

Parker \citep{Parker1955a,Parker1974a,Parker1974b,Parker1974c,Parker1979b} 
has pointed out that the magnetic field can be compressed to the enormous
intensity only by reducing the gas pressure within the flux tube relative to
the pressure outside, so that the external pressure compresses the field.
The only known mechanism for reducing the internal pressure sufficiently is a
reduction of the internal temperature over several scale heights so that the
gravitational field of the Sun pulls the gas down out of the tube (as described
by the known barometric law $dp / dz = - \rho g$). Hence it appears that the
intense magnetic field of the sunspot is a direct consequence of the observed
reduced temperature~\citep{Parker1955a}.

On the other hand, Parker \citep{Parker1974c,Parker1977} has also pointed out
that the magnetic inhibition of convective heat transport beneath the sunspot,
with the associated heat accumulation below, raises the temperature in the
lower part of the field. The barometric equilibrium leads to enhanced gas
pressure upward along the magnetic field, causing the field to disperse rather
than intensify. Consequently, \cite{Parker1974c} argued that the
temperature of the gas must be influenced by something more than the inhibition
of heat transport!

Our unique alternative idea is that the explanation of sunspots is
based not only on the suppression of convective heat transfer by a strong
magnetic field~\citep{Biermann1941} through the enhanced 
cooling of the Parker-Biermann effect~\citep{Parker1974a},
but also on the appearance of the axions of photonic origin
(Fig~\mbox{\ref{fig-lampochka}}, Fig.~B.1 in \cite{RusovDarkUniverse2021}) 
from the tachocline to the photosphere, which is confirmed by
the ``disappearance'' of the heat, and consequently, the temperature in the
lower part of the magnetic tube~\citep{Parker1974c,Parker1977} due to the axions
of photonic origin from the photon-axion oscillations in the O-loop near the
tachocline (see Fig.~\mbox{\ref{fig-lampochka}}).

This means that the appearance of axions of photonic origin,
which ``remove'' the problem of the temperature rise in the lower part of the
magnetic tube, and the photons of axionic origin, which have the free path
(Rosseland length; see Fig.~B.3 in~\cite{RusovDarkUniverse2021}) from the
tachocline to the photosphere, are the explanation of sunspots based not only
on the suppression of convective heat transport by a strong magnetic
field~\citep{Biermann1941}, but also on the indispensable existence of the
Parker-Biermann cooling effect.

On the other hand, we understand that the existence of the Parker-Biermann
cooling effect is associated with the so-called thermomagnetic
Ettingshausen-Nernst  effect (see Apendix~A in \cite{RusovDarkUniverse2021}).

Due to the large temperature gradient in the tachocline, the thermomagnetic EN
effect \citep{Ettingshausen1886,Sondheimer1948,Spitzer1956,Kim1969} creates
electric currents that are inversely proportional to the strong magnetic field
of the tachocline. As we showed earlier~\citep{RusovDarkUniverse2021}, the
toroidal magnetic field of tachocline by means of the thermomagnetic
Ettingshausen-Nernst effect (Appendix~A) ``neutralizes'' the magnetic field of
the solar core $\sim 5 \cdot 10^7~G$ \citep{Fowler1955,Couvidat2003}. It means
that, using the thermomagnetic EN effect, a simple estimate of the magnetic
pressure of an ideal gas in the tachocline of e.g. the Sun,

\begin{equation}
\frac{B_{tacho}^2}{8 \pi} = p_{ext} \approx 6.5 \cdot 10^{13} \frac{erg}{cm^3} ~~
at ~~ 0.7 R_{Sun}, 
\label{eq05-001}
\end{equation}

\noindent
can indirectly prove that by using the holographic principle of quantum gravity
(see Apendix~C in \cite{RusovDarkUniverse2021}), the repelling toroidal
magnetic field of the tachocline exactly ``neutralizes'' the magnetic field in
the Sun's core (see Fig.~A1 in~\cite{RusovDarkUniverse2021})

\begin{equation}
B_{tacho}^{Sun} = 4.1 \cdot 10^7 ~G = -B_{core}^{Sun} ,
\label{eq05-002}
\end{equation}

\noindent
where the projections of the magnetic fields of the tachocline and the core
have equal values but  opposite directions.

Such strong magnetic fields in a tachocline of e.g. the Earth, Sun, magnetic
white dwarfs, accreting neutron stars and BHs can predict the exact
``neutralization'' of the magnetic field in the core of these stars and in a
black hole (see Appendix~C in \cite{RusovDarkUniverse2021}).

Let us note another important Babcock-Leighton (BL) mechanism. On the one hand,
we mark out the holographic BL mechanism (see Fig.~C.1b in 
\cite{RusovDarkUniverse2021}), which we often refer to as the holographic
antidynamo mechanism, caused by a remarkable example of the Cowling antidynamo
theorem. This theorem states that no axisymmetric magnetic field can be
maintained through the self-sustaining action of the dynamo by means of an
axially symmetric current~\citep{Cowling1933}. On the other hand, the
holographic BL~mechanism (as a component of our solar anti-dynamo model)
follows our example of a thermomagnetic EN~effect, or the so-called solar
holographic antidynamo, in which the poloidal field originates directly from
the toroidal field, but not vice versa (see Fig.~C.1a in \cite{RusovDarkUniverse2021}).

%\begin{figure*}[tbp!]
\begin{figure*}
\noindent
\begin{center}
\includegraphics[width=\linewidth]{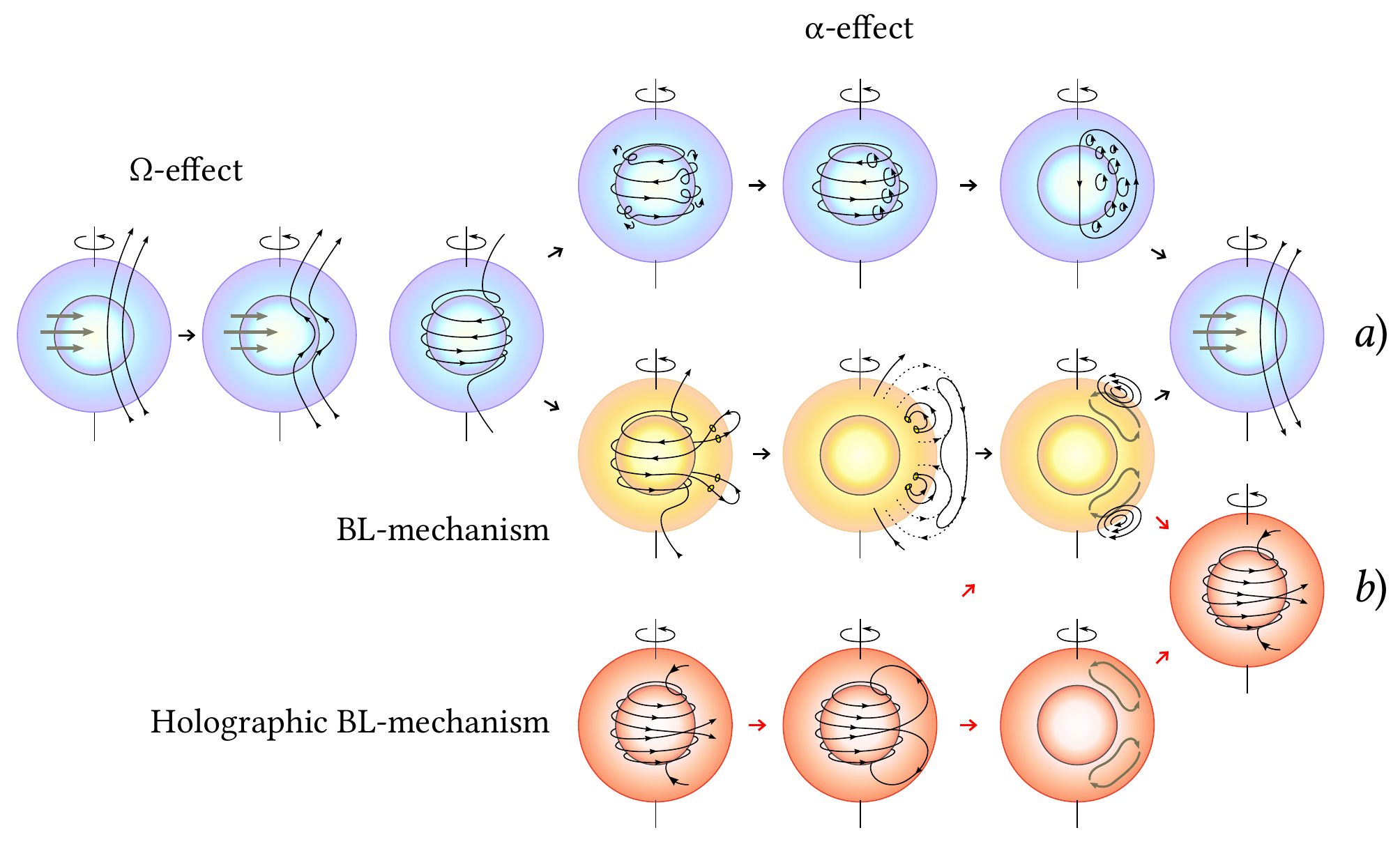}
\end{center}
\caption{An illustration of the main possible processes of a magnetically
active star of the Sun type.
(a) $\alpha$-effect, $\Omega$-effect and BL~mechanism as components of the
solar dynamo model. The $\Omega$-effect (blue) depicts the transformation of the
primary poloidal field into a toroidal field by differential rotation.
Regeneration of the poloidal field is then performed either by the
$\alpha$-effect (top) or by the BL~mechanism (yellow in the
middle). In case of $\alpha$-effect, the toroidal field at the base of the
convection zone is subject to cyclonic turbulence. In the BL~mechanism, the
main process of regeneration of the poloidal field (based on the 
$\Omega$-effect (blue)) is the formation of sunspots on the surface of the Sun
from the rise of floating toroidal flux tubes from the base of the convection
zone. The magnetic fields of these sunspots closest to the equator in each
hemisphere diffuse and join, and the field due to the spots closer to the
poles has a polarity opposite to the current that initiates rotation of the
polarity. The newly formed polar magnetic flux is transported by the meridional
flow to deeper layers of the convection zone, thereby creating a new
large-scale poloidal field. Derived from \cite{Sanchez2014}.
(b) BL~mechanism and holographic BL~mechanism as components of our
solar antidynamo model. Unlike the component of the solar dynamo model (a), the
BL~mechanism, which is predetermined by the fundamental holographic principle
of quantum gravity, and consequently, the formation of the thermomagnetic
EN~effect (see \cite{Spitzer1962,Spitzer2006,Rusov2015}),
emphasizes that this process is associated with the continuous transformation
of toroidal magnetic energy into poloidal magnetic energy
($T \rightarrow P$ transformation), but not vice versa ($P \rightarrow T$). This
means that the holographic
BL~mechanism is the main process of regeneration of the primary toroidal field
in the tachocline, and thus, the formation of floating toroidal magnetic flux
tubes at the base of the convective zone, which then rise to the surface of the
Sun. The joint connection between the poloidal and toroidal magnetic fields is
the result of the formation of the so-called meridional magnetic field, which
goes to the pole in the near-surface layer, and to the equator at the base of
the convection zone.}
%the convection zone. Adopted from~\cite{RusovDarkUniverse2021}}
\label{fig-solar-dynamos}
\end{figure*}

Using the Babcock-Leighton mechanism (see Fig.~\mbox{\ref{fig-solar-dynamos}}),
and consequently, the MFT in strong fields, we are interested in the existence
of dark matter axions identical to solar axions, which strongly affects the
magnetic O-loops inside the MFT near the tachocline, and is thus connected to
the so-called thermomagnetic Ettingshausen-Nernst effect through the
Parker-Biermann cooling effect (see Apendix~A in~\cite{RusovDarkUniverse2021}).

Since the photons of axion origin cause the Sun luminosity variations, then
unlike in the case of the self-excited dynamo, an unexpected yet simple
question arises: is there a dark matter chronometer hidden deep in the Sun
core?

In order to answer this question, let us first consider all the unexpected and
intriguing implications of the 11-year modulations of the ADM density in the
solar interior and around the BH (see Sect.~3 in~\cite{RusovDarkUniverse2021}).

A unique result of our model is the fact (see 
Sect.~3 in~\cite{RusovDarkUniverse2021}), that the periods, velocities
and modulations of the S-stars are the essential indicator of the modulation
of the ADM halo density in the fundamental plane of the Galaxy center, which
closely correlates with the density modulation of the baryon matter near the
SMBH. If the modulations of the ADM halo at the GC lead to modulations of the
ADM density on the surface of the Sun (through vertical density waves from the
disk to the solar neighborhood), then there is an "experimental"
anticorrelation identity between such indicators as the ADM density modulation
in the solar interior and the number of sunspots. Or equivalently, between the
modulation of solar axions (or photons of axion origin) and the sunspot
cycles! 

A hypothetical pseudoscalar particle called axion is predicted by the theory
related to solving the CP-invariance violation problem in QCD. The most
important parameter determining the axion properties is the energy scale $f_a$
of the so-called U(1) Peccei-Quinn symmetry violation. It determines both the
axion mass and the strength of its coupling to fermions and gauge bosons
including photons. However, in spite of the numerous direct experiments, axions
have not been discovered so far. Meanwhile, these experiments together with the
astrophysical and cosmological limitations leave a rather narrow band for the
permissible parameters of invisible axion (e.g.
$10^{-6} eV \leqslant m_a \leqslant 10^{-2} eV$~\citep{ref01,ref02}).
% which is
%also a well-motivated cold dark matter (CDM) candidate in this mass region
%\cite{Preskill1983,Dine1983,Abbott1983}.
The PQ mechanism, solving the strong CP problem in a very elegant 
way~\citep{PecceiQuinn1977,PecceiQuinn1977PRD}, is
especially attractive here, since the axion is also a candidate for
dark matter~\citep{Preskill1983,Abbott1983,Dine1983,Kawasaki2013,Marsh2016,DiLuzio2020,Sikivie2021}.

The following question is very important for us: Where do relict hot dark
matter axions\footnote{
%Axions can be hot or cold dark matter axions (see e.g.
%the last paragraph of the Introduction in~\cite{Sikivie2021})
%Rev5?
See e.g. the works~\cite{Turner1987,Archidiacono2015,CAST2017,DiLuzio2021}.}
in cosmology come from?

%Rev4
In this regard, we will consider the remarkable model by Shafieloo, Hazra,
Sahni and Starobinsky~\cite{Shafieloo2017}, who proposed a new description of
the cosmological constant $\Lambda$ that determines the density of dark energy
(DE). A model in which (surprisingly enough!) metastable dark energy decays
radioactively into dark matter, leads to lower values of the Hubble parameter
at high redshifts compared to cold dark matter (CDM) in a form of the final
stage of the inflation. The analogue of the model by Shafieloo, Hazra,
Sahni and Starobinsky~\cite{Shafieloo2017}:

%Rev4
\begin{equation}
\Omega_{DE} \rightarrow \Omega_{DM} = \Omega_{CDM} + \Omega_{baryon} .
\end{equation}

%Rev4
Although this model provides a better fit to cosmological data of baryon
acoustic oscillation (BAO) (especially data from high-redshift quasars) than
concordance ($\Lambda$CDM) cosmology, we are also interested in real physics of
dark matter.

\begin{figure}
\noindent
  \begin{center}
    \includegraphics[width=10cm]{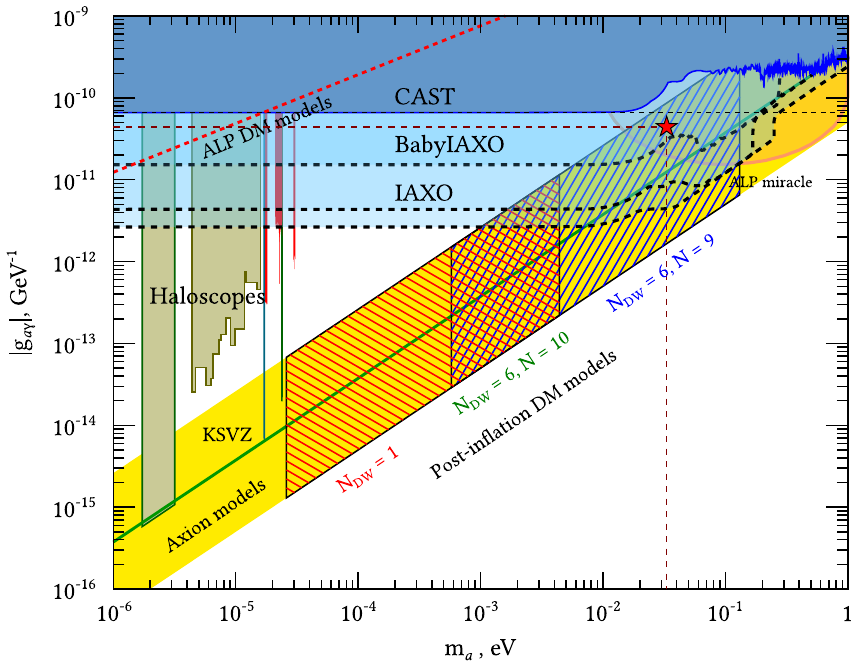}
  \end{center}
\caption{Predicted mass ranges in which QCD axions can account for the
observed cold DM abundance in the post-inflationary PQ symmetry breaking
scenario. The generic prediction of QCD axion models is shown in the yellow
region. The sensitivity prospects of IAXO are also plotted. The prediction of
the post-inflationary PQ symmetry breaking scenario differs according to the
value of $N_{DW}$ and the structure of the explicit symmetry breaking terms in
the models with $N_{DW} > 1$: $N_{DW}$ is the colour-anomaly of the PQ
symmetry, which turns out to be a positive integer and it is called the 
``domain wall number'' in this context. Dashed regions correspond to the
predictions (up to uncertainties in the estimation of the relic axion
abundance) of the models with $N_{DW} = 1$~(red), $N_{DW} = 6$ and 
$N = 9$~(blue), and $N_{DW} = 6$ and $N = 10$~(green). Adopted 
from~\cite{Armengaud2019}. The vertical green lines are 
from~\cite{Backes2021} and~\cite{Brubaker2017}, three red lines
are from~\cite{Boutan2018}. The red star marks the axion mass 
$m_a \sim 3.2 \cdot 10^{-2}~eV$ and the axion-photon coupling constant
$g_{a\gamma} \sim 4.4\cdot 10^{-11}~GeV^{-1}$, obtained 
in~\cite{RusovDarkUniverse2021} based coronal heating problem solution
by means of axion origin photons and the ADM.}
\label{fig-axion-constraints}
\end{figure}

\begin{figure}
  \begin{center}
  \includegraphics[width=15cm]{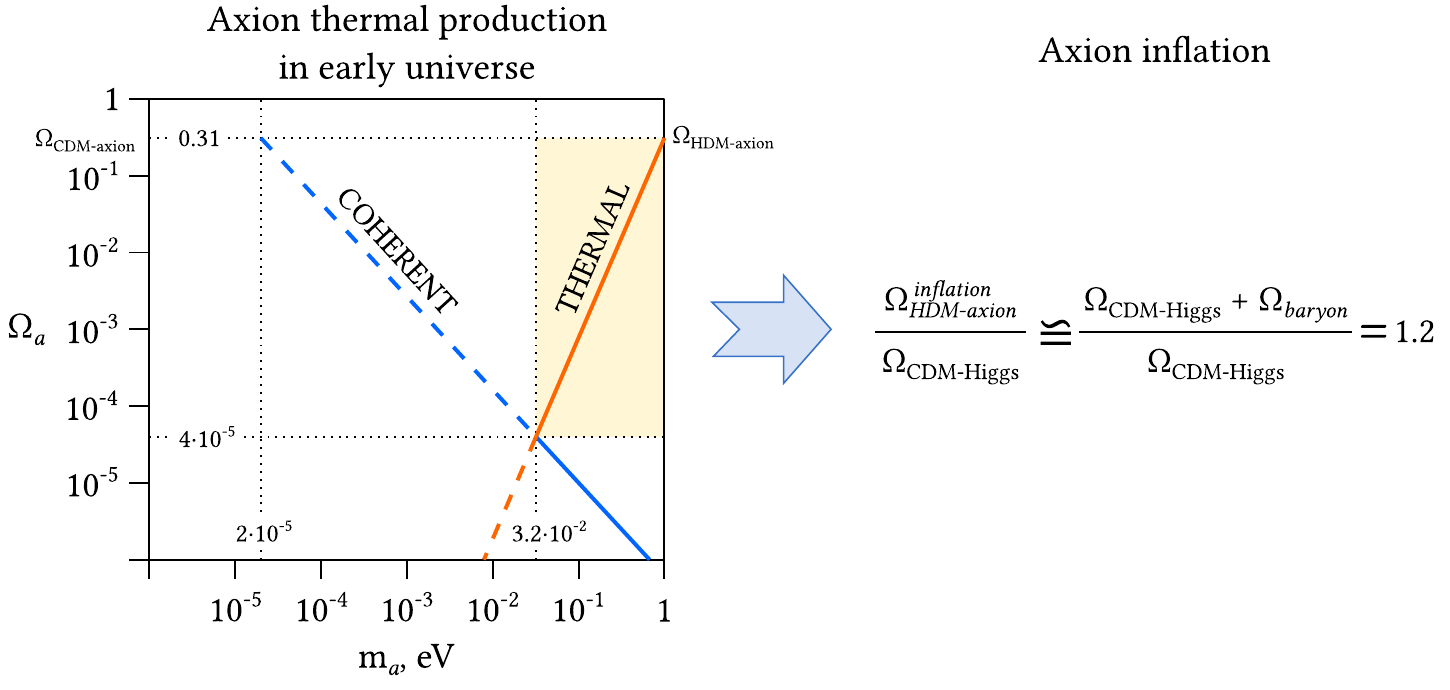}
  \end{center}
\caption{Thermal and coherent axion contributions to the fraction of critical
density ($\rho_c = 3 H_0^2 / 8\pi G$) as a function of axion mass (left panel
derived from~\cite{Turner1987}). On the one hand, 
$m_a \sim 3.2\cdot 10^{-2}~eV$ sets the identity of the thermal and cogerent
axions ($\Omega_{thermal} / \Omega_{coherent} \sim 1$), and on the other hand,
for axion masses $\sim 1~eV$, the coherently produced axions are thermalized,
so only for $m_a \leqslant 1~eV$ (see~\cite{PDG2020}) there are two axion
populations -- thermal (red) and coherent (blue). As a result (see the right
panel), the axion inflation in the Universe leads to baryon 
asymmetry~\citep{Co2020,Co2021}, and also induces the charge asymmetry for the
Higgs boson cold dark matter (see~\cite{Merkotan2018,Ptashynskiy2019}) with a
rather large mass of $\sim$5~GeV (see Fig.~6 in~\cite{Vincent2016}), which at
the same time forms the CDM Higgs halo (right panel).}
\label{fig-axion-inflation}
\end{figure}

We know that, according to the excellent work~\cite{Turner1987}, the 
Peccei-Quinn symmetry breaking scale is calculated to be $\sim 2\cdot 10^8~GeV$
(corresponding to the axion mass of $\sim 3.2 \cdot 10^{-2} ~eV$ (see 
Fig.~\ref{fig-axion-constraints}, and also Fig.~1 
in~\cite{RusovDarkUniverse2021})), at which the thermal production of hot dark
matter axions in the early Universe (through Primakoff processes 
($q  + \gamma \rightarrow q + a$; the particle in the loop is the heavy $Q$
quark) and photoproduction on heavy quarks (see Fig.~1 in~\cite{Turner1987}))
dominates over coherent production of dark matter axions as approximately 
$\Omega_{thermal} / \Omega_{coherent} \sim 1200 \times [m_a / 1eV]^{2.175}$,
where $\Omega_{coherent}$ was calculated in~\cite{Turner1986,Preskill1983,
Abbott1983,Dine1983,Kawasaki2013,Marsh2016,DiLuzio2020,Sikivie2021}. The photon
luminosity from the decays of these relict axions leads to an upper limit of
the axion mass of the order of $\sim 1~eV$ (see left panel in 
Fig.~\ref{fig-axion-inflation}). If the mass of the axion saturates this bound,
then it is quite possible to detect the relict axion decay.

%\begin{figure}[tb!]

%In other words, as soon as the QCD scale becomes small enough, the axion
%begins to rotate, and the kinetic energy of rotation becomes greater than the
%barrier of the axion's cosine potential (see Fig.~\mbox{\ref{fig-rotating-axion}}).
%As a result, the number of axions increases~\citep{Co2020}, thus generating
%a high density of ``invisible'' axions in the inflationary Universe (see 
%e.g.~\cite{EingornRusov2015}).
%
%%Другими словами, как только масштаб КХД становится достаточно малым, аксион начинает вращаться, причем кинетическая энергия вращения становится больше, чем ба-рьер косинусоидального потенциала аксиона (see Fig.3). Как результат, количество ак-сионов увеличивается Co  Harigaya 2020 и, тем самым, порождяют высокую плотность “невидимых” аксионов в инфляционной Вселенной see. e,g., Eingorn  Rusov 2015.

%Rev5
But the most remarkable fact is that the thermal production of axions by hot
dark matter in the early Universe, where inflation occurs before PQ-symmetry
breaking, happens through the increasing rate of axion production and leads to
a large number of axions with masses greatly exceeding $3.2 \cdot 10^{-2}~eV$, 
i.e. about $1~eV$. Therefore, the thermally produced axions (with 
$\Omega_{HDM-axion} \sim 0.31$, see left panel in 
Fig.~\ref{fig-axion-inflation}) dominate the population of relic axions, and
during the inflation, due to the energy of the axion field, the hot dark matter
axions are converted to other fields and generate particles. The latter
manifest themselves as the excess of baryons ($\Omega_{baryon} \sim 0.049$)
over antibaryons (see e.g.~\cite{Domcke2020}), and also in the form of 
asymmetric CDM Higgs\footnote{In contrast to the (incomplete) Standard Model, the Higgs-boson here is supposed to have integer weak isospin, which is consistent with the experimental data on its decay channels~\citep{PDG2020}. It is considered as a bound state of $W^+$ and $W^-$ bosons -- the particles with weak isospin of 1. Since the weak isospin of gauge bosons is equal to unity, the bound state can have a weak isospin of 2, 1 or 0. The state with zero isospin obviously cannot take part in weak interactions. It has zero electric charge and does not consist of strongly inteacting particles. Therefore, such a state can interact only gravitationally. On the other hand, it has all the theoretical properties of the Higgs boson. For example, in the model of multiparticle fields, the spontaneous symmetry breaking is NOT postulated as in the Standard Model (see~\cite{Merkotan2017,Merkotan2018,Ptashynskiy2019}), but is obtained as a result of the dynamic equations of the model, as a result of which it has a nonzero vacuum value. Due to these properties, it can be considered as a candidate for dark matter~\citep{Merkotan2017,Merkotan2018,Ptashynskiy2019}.} (see right panel in Fig.~\ref{fig-axion-inflation}) and the formation of some remnant thermalized axions with mass $3.2 \cdot 10^{-2}~eV$ involved in the CDM Higgs halo.
We also know~\cite{RusovDarkUniverse2021} that the asymmetry of the CDM Higgs boson~\cite{Merkotan2017,Merkotan2018,Ptashynskiy2019} with a mass of $\sim$5~GeV (see Fig.~6 in~\cite{Vincent2016})
and a high density $\Omega_{CDM-Higgs} \sim 0.26$ forms in the final stage of inflationary Universe and exists till nowadays.

%Rev5

The main result is that the axions of hot dark matter (HDM axion; see the left
panel in Fig.~\ref{fig-axion-inflation}) predetermine the formation of baryon
asymmetry and asymmetry of the cold dark matter Higgs boson (CDM-Higgs; 
see~\cite{Merkotan2017,Merkotan2018,Ptashynskiy2019}), as well as the formation
of some remnants of hot dark matter axions (HDM axion) participating in the
CDM Higgs halo (see analog of~\cite{RusovDarkUniverse2021}). This leads to
the following description of the generalized model (which is directly related
to the Starobinsky model (see $\Omega_{DE}$ in~\cite{Shafieloo2017}), 
%the Co~\&~Harigaya model (see $\Omega_{HDM-axion}$ in~\cite{Co2020})
the Turner model (see $\Omega_{HDM-axion}$
in~\cite{Turner1987}) and the Sharph~\&~Rusov model (see $\Omega_{CDM-Higgs}$
in~\cite{Merkotan2017,Merkotan2018,Ptashynskiy2019}) in the form 

\begin{align}
\Omega_{DE} \rightarrow \Omega_{DM} \equiv  \Omega_{HDM-axion} & = \Omega_{CDM-Higgs} + \Omega_{baryon} + \Omega_{(solar)axion} \nonumber \\
& \cong \Omega_{CDM-Higgs} + \Omega_{baryon}
\label{eq-5.1-R4-02}
\end{align}
\noindent where 

\begin{center}
\hfill $\Omega_{HDM-axion} \equiv \Omega_{HDM-axion}^{thermal} (m_a \sim 1eV) + {}$ \hfill ~ \newline
${} + \Omega_{HDM-axion}^{thermal} (m_a \sim 3.2 \cdot 10^{-2}eV) + \Omega_{HDM-axion}^{coherent} (m_a \sim 3.2 \cdot 10^{-2}eV) = $ \newline
$ = \Omega_{HDM-axion}^{thermal} (m_a \sim 1eV) + \Omega_{(solar)axion} (m_a \sim 3.2 \cdot 10^{-2}eV) \cong$ \newline
$\cong \Omega_{HDM-axion}^{thermal} (m_a \sim 1eV) = 0.31$
\end{center}
(see left panel in Fig.~\ref{fig-axion-inflation}) when 
$\Omega_{CDM-Higgs} > \Omega_{baryon} \gg \Omega_{(solar)axion}$ (see Eq.~\ref{eq-5.2-09}).

\begin{figure}
\begin{center}
  \includegraphics[width=16cm]{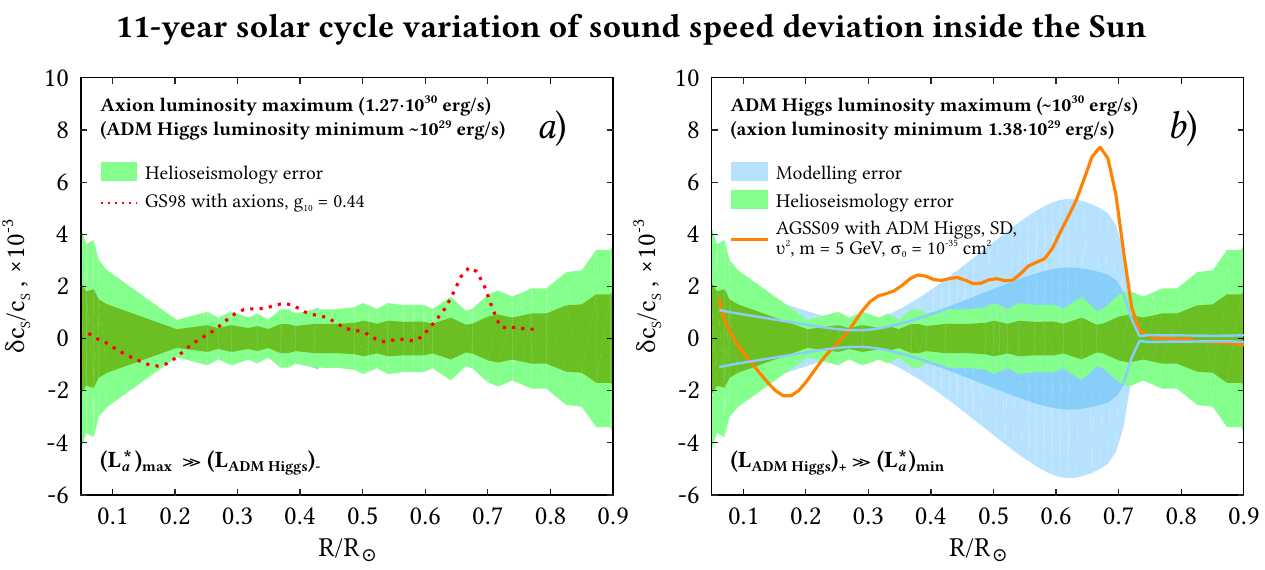}
\end{center}
\caption{Deviation of the radial speed of sound profile (Sun - model)/Sun in
the solar interior from values obtained for two models: (a) the red dots from
the values of \cite{Grevesse1998} with axions almost coincide with the values 
in~\cite{Vinyoles2015}
and (b) the asymmetric Higgs dark matter captured inside the Sun (solid orange
line of ADM~\citep{Vincent2016}).
%the Higgs
%asymmetric dark matter (solid red line ADM;~\cite{Vincent2016}) and the solar axions (red
%dotted lines;~\cite{Grevesse1998}).
Colored areas indicate 1$\sigma$ and 2$\sigma$ errors in modeling (thick blue
bar;~\cite{Vincent2015a,Vincent2015b}) and helioseismological inverses (thinner
green bar;~\cite{DeglInnoccenti1997,Fiorentini2001}).
The deviation of the radial speed of sound is associated with the part of the
corona luminosity~\citep{RusovDarkUniverse2021} in the total luminosity of the
Sun, which can be easily determined from (a) the solar maximum (involving the
maximum axion luminosity fraction ($(L_a^*)_{max} \gg (L_{ADM Higgs})_{-}$) and
(b) the solar minimum (involving minimum fraction of the axion luminosity 
($(L_a^*)_{min} \ll (L_{ADM Higgs})_{+}$) and is in good agreement with
nonstandard solar model.}
\label{fig-helioseismology}
\end{figure}

The latter, surprisingly enough, are the ``experimental'' key to the existence
of  the asymmetry of CDM Higgs boson and the ``remnants'' of cold dark matter
axions (see Eqs.~(\ref{eq-5.1-R4-02}) and~(\ref{eq-5.2-09})) -- this is a
direct proof of the solution to the problem of solar corona heating, which is
associated with the hot dark matter axions on the Sun ($\Omega_{(solar)axion}$
in~(\ref{eq-5.1-R4-02})) and the asymmetry of the CDM Higgs boson halo in solar
neighborhood\footnote{In this case, neutrinos and helioseismic information thus
complement each other, especially when the presence of particles such as axions
at the maximum luminosity of the Sun, and particles such as the Higgs ADM at
the minimum luminosity of the Sun, affects the transfer of heat within the
interior of the Sun.
Therefore, the maximum and minimum of neutrino fluxes can simultaneously solve
a very complex ``problem of solar composition'', or otherwise called ``the
problem of solar abundance'', which is associated with the solar cycle.
Most important here is that the neutrino flux from the subdominant CNO cycle is
linearly dependent on the metallicity ($Z$) of the solar core (see 
e.g.~\cite{Gann2015,Borexino2020a,Borexino2020b,TapiaArellano2021}). 
It means that the modulation of solar abundance, with high metallicity at the
solar maximum ($Z\sim 0.0170$ with $Z/X=0.023$~\citep{Grevesse1998}) and
low metallicity at the solar minimum ($Z\sim 0.0133$ with 
$Z/X=0.0178$~\citep{Asplund2009}), is in good agreement with the
deviation of the radial sound speed profile 
$\delta c_s / c_s = (c_{s,obs} - c_{s,th})/c_{s,obs}$ inside the Sun 
(Fig.~\mbox{\ref{fig-helioseismology}}), which consists of two related models:
for the maximum luminosity of axions (Fig.~\mbox{\ref{fig-helioseismology}}a)
and for the minimum luminosity of axions, which is associated with the maximum
luminosity of Higgs ADM (Fig.~\mbox{\ref{fig-helioseismology}}b).}
and which is anti-correlated (through vertical density
waves from the disk to the Sun) with 11-year S-stars around the black hole (see
Fig.~\ref{fig-helioseismology} and Eqs.~(27)-(28) 
in~\cite{RusovDarkUniverse2021}).
Second, let us remind that the most fundamental problem of the Standard Model
is the appearance of the scalar component of the two-particle gauge field in a
form of the Higgs boson, scalar in interior indices, which is considered a
component of dark matter (see Fig.~\ref{fig-helioseismology})!

%\begin{figure}[tb!]

So what is so special about this Higgs dark matter particle? We know that the
existence of dark matter in the Universe is a striking evidence that physics
goes beyond the Standard Model, although its nature remains a mystery. At the
same time, the closeness of the energy densities of DM and baryons 
$\Omega_{DM} \approx 5.3 \Omega_{baryon}$~\citep{PlanckCol2015,Patrignani2016}
motivates the idea of asymmetric dark matter (ADM)~\citep{Nussinov1985,
Chivukula1990, Barr1990,Kaplan1992,Hooper2005,Kaplan2009,Davoudiasl2012,
Petraki2013,Zurek2014},
based on the assumption that the current DM density is determined by the
$\eta_{DM}$ asymmetry in the DM sector, similar to the baryon asymmetry
$\eta_{baryon}$. Then

\begin{equation}
\frac{\Omega_{DM}}{\Omega_{baryon}} = \frac{m_{DM}}{m_{N}} 
\frac{\eta_{DM}}{\eta_{baryon}}
\label{eq-5.2-06}
\end{equation}

\noindent
where $m_{DM}$ is the DM mass and $m_N$ is the nucleon mass. If two
asymmetries are generated by the same mechanism, or one dark asymmetry is
responsible for the other -- baryon asymmetry arising from leptogenesis, we
expect $\eta_{DM} \sim \eta_{B}$. Hence it follows that taking the value 
$m_{DM} = m_{ADM} \sim 5~GeV$ and $m_N \approx 0.939~GeV$, one can obtain
the value

\begin{equation}
\frac{\Omega_{DM}}{\Omega_{baryon}} = \frac{m_{DM}}{m_{N}} 
\frac{\eta_{DM}}{\eta_{baryon}} \approx
\frac{m_{ADM}}{m_{N}}  \approx 5.32 ,
\label{eq-5.2-07}
\end{equation}

\noindent
which suggests that, according to~\cite{Turner1987}, 
$\Omega_{HDM-axion} \sim 0.31$ (see Fig.~\ref{fig-axion-inflation} and
Eq.~(\ref{eq-5.1-R4-02})) coincides with the high density (thermal!)
hot dark matter axions (see Fig.~\ref{fig-axion-inflation}) during the
Universe inflation (see right panel in 
Fig.~\ref{fig-axion-inflation}).

%Rev4
Applying the equality $\Omega_{DM} = \Omega_{ADM}$, we use, according to Eq.~(\ref{eq-5.1-R4-02}), the identity between $\Omega_{ADM}$ and $\Omega_{CDM-Higgs}$, which allows one to obtain the following equation

\begin{align}
& \frac{\Omega_{HDM-axion}^{inflation}}{\Omega_{CDM-Higgs}} =
\frac{\Omega_{CDM-Higgs} + \Omega_{baryon}  + \Omega_{(solar)axion}}{\Omega_{CDM-Higgs}} \cong \nonumber \\
& \frac{\Omega_{CDM-Higgs} + \Omega_{CDM-Higgs} / 5.32}{\Omega_{CDM-Higgs}}
= 1.2 ,
\label{eq-5.2-08}
\end{align}

\noindent
where the new physics manifests not only Eq.~(\mbox{\ref{eq-5.2-08}}) (see
right panel in Fig.~\mbox{\ref{fig-axion-inflation}}), but also a description
of the low density of hot dark matter axions (see Eq.~(17) 
in~\cite{Graham2015}) in the final stage of the Universe inflation and till
nowadays.

\begin{equation}
\Omega_{(solar)axion} \approx
2 \left( \frac{6~\mu eV}{m_a} \right)^{7/6} \approx 10^{-4} ,
\label{eq-5.2-09}
\end{equation}

\noindent
%which is experimentally obtained in the form of the estimated hadronic
%axion-photon coupling and mass $m_a$ (see 
%Eq.~(\mbox{\ref{eq-axion-parameters}}), Fig.~\mbox{\ref{fig-axion-constraints}}
%and Fig.~1 in~\mbox{\cite{RusovDarkUniverse2021}}).
which, on the basis of ``experimental'' data, identically coincides with the
values of the hadronic axion-photon coupling 
$g_{a\gamma} \sim 4.4 \cdot 10^{-11} ~GeV^{-1}$ and the mass 
$m_a \sim 3.2 \cdot 10^{-2}~eV$ on the Sun (see 
Fig.~\ref{fig-axion-constraints} and Fig.~1 
in~\cite{RusovDarkUniverse2021}).
This means that the axions of hot dark matter on the Sun and the axions of hot
dark matter in our galactic Higgs ADM halo (see  Eq.~(\ref{eq-5.2-09})) are
absolutely identical!

%Most importantly, this means that the same axions can be the hot dark matter,
%e.g. on the Sun, or the axions of CDM in the Galaxy, although the axion
%component is very insignificant (see Eq.~\mbox{(\ref{eq-5.2-09})}) in the ADM
%halo~\citep{RusovDarkUniverse2021}.
%As a consequence, the flux of hot dark matter axions with high FSL, emitted
%from the core through the surface of the Sun, can scatter across the galaxy
%without high FSL, i.e. goes to the redshift of nonrelativistic energies, and
%surprisingly, with time, indentically coinciding with the
%axions of CDM in the ADM halo from the center to the edge of the galaxy.
%
%%Самое главное, это означает, что одни и те же аксионы темной материи могут быть горячей темной материей, например, на Солнце, или, например, аксионами холодной тем-ной материи в Галактике, хотя аксионная компонента очень незначительна (see Eq.8) в гало ADM (see Rusov et al. 2021). Как следствие, поток аксионов горячей темной материи с высоким FSL, излучаемых из ядра через поверхность Солнца, могут рассеиваться по галактике вовсе не на высоких FSL, с течением времени, как ни странно, идентично  совпадая с аксионами холодной темной материей  в гало ADM от центр к краю галактики.

This is due to the fact that the PQ symmetry is significantly broken in the
early Universe, and is caused by the production of dark matter axions. The most
surprising is the fact that the thermal production of the axion of hot dark
matter in early Universe (see left panel in Fig.~\ref{fig-axion-inflation}) is
an inflationary source of the formation of asymmetry of baryons, in which,
during post-inflation, the ``collisions and interactions between gas-rich
galaxies are considered key stages in their formation and evolution, causing
the rapid formation of new stars''~\cite{Goulding2018}, and, as a consequence,
emitting streams (!) of hot dark matter axions (see e.g. 
supernovae~\cite{Hsueh2019}, Galactic Globular Clusters~\cite{Gilman2019}, and
the Sun~\cite{RusovDarkUniverse2021}).

But the most remarkable fact is that our data on the axions dark matter in
the form of the estimates of the hadronic axion-photon coupling $g_{a\gamma}$,
the mass $m_a$ and the ADM Higgs mass, experimentally coincide, according 
to~\cite{RusovDarkUniverse2021}, with the solution for the ADM Higgs
chronometer in the S-stars interior, which controls the solar cycles of the
dark matter axions luminosity, and consequently, is the remarkable solution for
the problem of solar corona heating using photons of axion origin and ADM Higgs
-- a complete proof of the existence of the ADM Higgs~\cite{RusovDarkUniverse2021}.

And, finally, the fundamental results in both directions make it possible to
obtain (through indirect detection) the convincing evidence of the existence
and nature of the asymmetry of cold dark matter Higgs halo near a black hole 
and, as a result, in the solar neighborhood.
They also demonstrate a remarkable identity of solar axions and axions of hot
dark matter in the Universe (see $m_a \sim 3.2 \cdot 10^{-2}~eV$ of the thermal
and cogerent axions in the left panel of Fig.~\ref{fig-axion-inflation}). 
At the same time, it must be remembered that initially an axion with a mass of
1~eV, which is a low-energy remnant of the Peccei-Quinn solution to the strong
CP problem~\cite{PecceiQuinn1977,PecceiQuinn1977PRD,Wilczek1978,Weinberg1978},
makes the main contribution to the energy density of the Universe (see 
$\Omega_{HDM-axion} \sim 0.31$ in Fig.~\ref{fig-axion-inflation} and 
Eq.~(\ref{eq-5.1-R4-02})). Although it is not accessible for direct
cosmological searches for hot dark matter axions, it is accessible for direct
astrophysical experiments for axions of hot dark matter with a mass of 
$m_a \sim 3.2 \cdot 10^{-2}~eV$, identical to solar axions!

%The clue that dark matter axions do exist is therefore not only the evidence of
%e.g. the duration of the SN1987A neutrino burst, astrophysical boundaries from
%the horizontal branch and massive stars that are labeled
%HB~\cite{Raffelt1988SN1987A} and "Cepheids"~\cite{Carosi2013}, but also a
%surprising remarkable solution to the coronal heating
%problem~\cite{RusovDarkUniverse2021}! On the other hand, the estimates of axion
%dark matter are not yet violated by experimental data based on the well-known
%quantum enhanced search for dark matter axions (see e.g. 
%\cite{Bertone2018,Backes2021}).

Taking into account the above remarks, we arranged our paper in the following
way.
In Section~\ref{sec-empty-tubes} we discuss the effect of virtually empty
magnetic tubes and dark matter axions in the Sun's core. 
In Section~\ref{sec-radiative-heating} we present the convective heating and
buoyancy of magnetic tubes -- the phenomenon of solar axions of dark matter.
In Section~\ref{sec-tilt} we discuss the physics of magnetic reconnection of
the magnetic tubes in the lower layers and the observed features of the tilt
angle of Joy's law. 
Finally, in Section~\ref{sec-summary} we provide a summary and perspective of
this work.

\section{Effect of virtually empty magnetic flux tubes and  solar axion in the Sun interior}
\label{sec-empty-tubes}

The appearance of sunspots on the solar surface is one of the major
manifestations of solar activity demonstrating the cyclic behavior with the
period of about 11 years (see e.g. \cite{Hathaway2015}). A very high density
of the magnetic field is observed within the sunspots, which suppresses the
convective heat flow from the solar interior to the surface
\citep{Biermann1941,Alfven1942,Cowling1953,Parker1955a}. That is why the sunspots are cooler
and darker against the background of the solar disk. More than a hundred years
ago~\citep{Hale1908} discovered the vertical ``vortices'' of the magnetic field in
the sunspots. About a year later British astronomer~\citep{Evershed1909} was
conducting the observations in Kodaikanal (Tamil Nadu, India) and found that in
spite of the Hale's vertical ``vortices'', the magnetic field in the penumbra is
radially divergent from the center of a sunspot. The mechanism of the
sunspot formation including the umbra and penumbra as well as the Evershed
effect are still the subject of active discussions and studies today, and lots
of fundamental questions remain unanswered 
\citep{Solanki2003,Borrero2011,Tiwari2015,Pozuelo2015}. We consider some
possible solutions to these problems below.

We are mostly interested in effects for which the current theories, assuming
that the sunspots are produced by the dynamo action at the bottom of the
convection zone, fail to provide the convincing proofs and explain the dynamo
action in the convection zone. This problem becomes even stronger with the
recent findings. The numerical simulations of the solar dynamo have not
revealed thin MFTs of the comparable strength so far 
(see e.g. \cite{Guerrero2011,Nelson2013,Kapyla2013}). The helioseismology
does not give any evidence of the upward MFTs existence either
\citep{Birch2013,Birch2016}. This is especially important because of the
recent helioseismological investigations which set the strict limitations on
the velocities of large-scale convection inside the Sun and demonstrated
the inconsistency with the existing global magnetoconvection e.g.~\cite{ref34-3} modeling
\citep{Hanasoge2010,Hanasoge2012,Hanasoge2015,Gizon2012,Birch2013,Birch2016}.

Thus, our major question is: ``How are the sunspots generated by the strong
magnetic field at the base of the convection zone without any dynamo action?'',
or otherwise ``Which fundamental processes connect the sunspot cycle with the
large-scale magnetic field of the Sun?''. Or even more precisely, 
``What are the fundamental processes, associated with solar axions,
connect the sunspot cycle to the large-scale magnetic field of the Sun?''

%\begin{figure*}[tbp]
\begin{figure*}[tbp]
  \begin{center}
    \includegraphics[width=14cm]{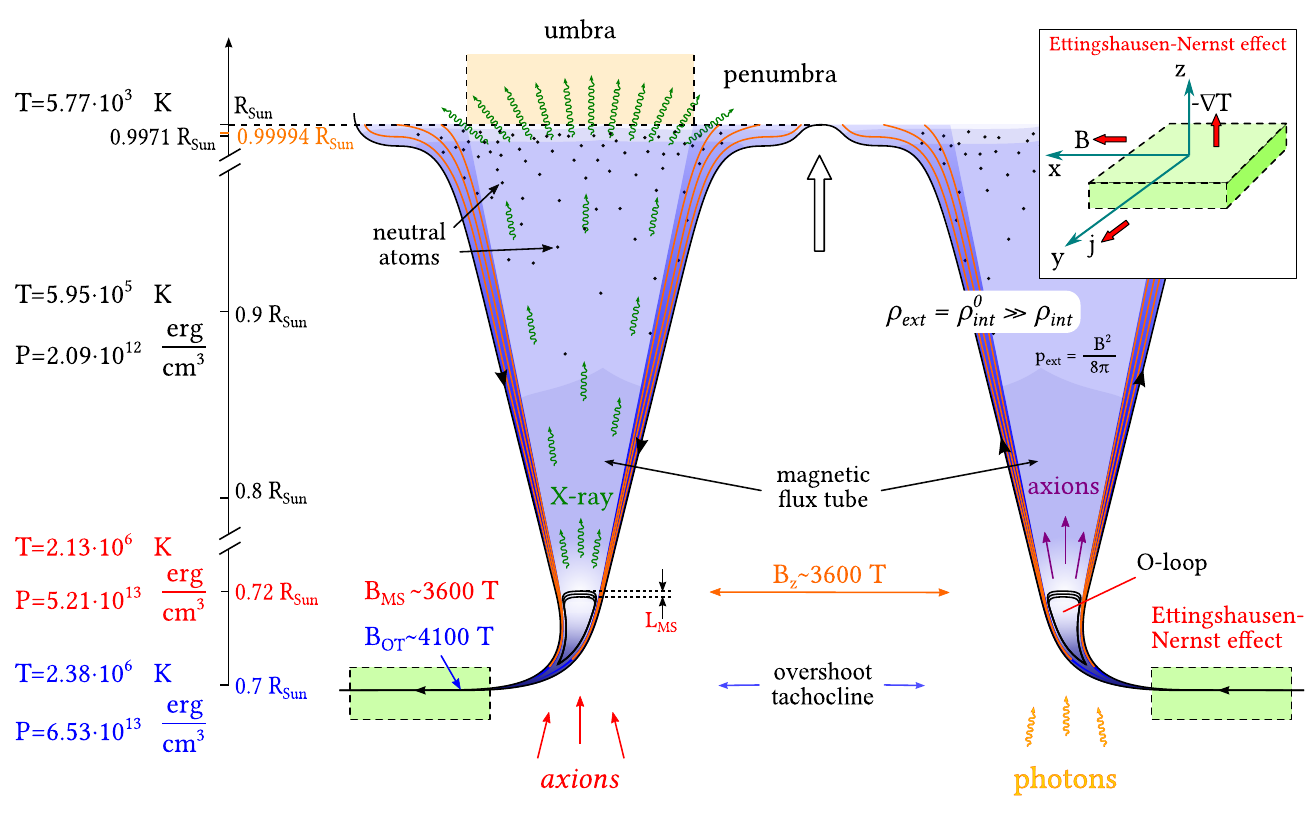}
  \end{center}
\caption{Topological effects of magnetic reconnection inside the
magnetic tubes with the ``magnetic steps'' (Fig.~B.1 in \cite{RusovDarkUniverse2021}). The left panel shows the
temperature and pressure change along the radius of the Sun from the tachocline
to the photosphere \citep{Bahcall1992}, $L_{MS}$ is the height of the magnetic
shear steps. At $R \sim 0.72~R_{Sun}$ the vertical magnetic field,
which is developed from the horizontal magnetic field through the well-known
Kolmogorov turbulent cascade (see Fig.~\ref{fig-Kolmogorov-cascade}),
reaches $B_z \sim 3600$~T, and the magnetic pressure $p_{ext} = B^2 / 8\pi 
\simeq 5.21 \cdot 10^{13}~erg/cm^3$ \citep{Bahcall1992}. The very cool regions 
along the entire convective zone caused by the Parker-Biermann cooling effect 
have the virtually zero internal gas pressure, i.e. the maximum magnetic pressure in the magnetic tubes.
}
\label{fig-lampochka}
\end{figure*}

Of all the known concepts playing a noticeable role in understanding of the
link between the energy transfer and the darkness of sunspots, let us
consider the most significant one, in our opinion. It is based on the
Parker-Biermann cooling effect \citep{Parker1955a,Biermann1941,Parker1979b} in
strong fields, which explains how the suppression of the Parker's convective 
heat transport in the lower part of the magnetic tube manifests itself.

In order to understand the physics of the Parker's suppression of the
convective heat transport in strong magnetic fields, we need to tun to the dark
matter axions, born in the core of the Sun. With solar axions
and the existence of a magnetic O-loop inside the MFT near the tachocline, the answer becomes very simple. When a magnetic O-loop is formed inside the MFT near the tachocline through the Kolmogorov turbulent cascade (see Fig.~2), the high-energy photons from the radiation zone experience the  axion-photon oscillations in this O-loop, and the so-called axions of photonic origin appear under the sunspot. This means that the cooling effect of Parker-Birmann exists due to the disappearance of barometric equilibrium~\citep{Parker1974a} and, as a consequence, the manifestation of the photon free path (Rosseland length; see Fig.~B.3 in ~\cite{RusovDarkUniverse2021}) from the tachocline to the photosphere, which is confirmed by axions of photonic origin after photon-axion oscillations in the O-loop (see Fig.~\ref{fig-lampochka}).

%\begin{figure}[tbp]
\begin{figure}
\begin{center}
\includegraphics[width=12cm]{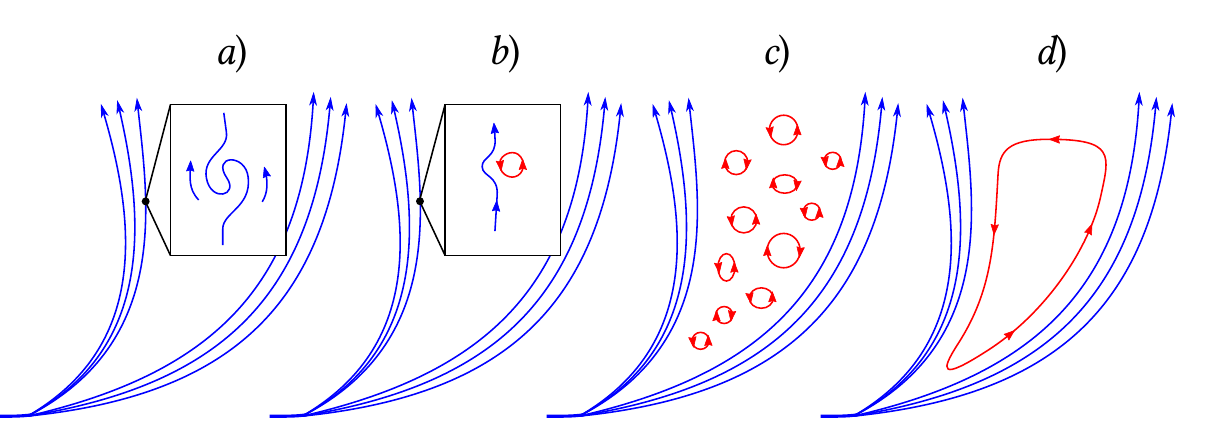}
\end{center}
\caption{Kolmogorov turbulent cascade \citep{Kolmogorov1941,Kolmogorov1968,Kolmogorov1991} and magnetic reconnection in the lower layers inside the unipolar magnetic tube (Fig.~B.2 in \cite{RusovDarkUniverse2021}). Common to these various turbulent systems is the presence of the inertial range of Kolmogorov, through which the energy is cascaded from large to small scales, where dissipative mechanisms (as a consequence of magnetic reconnection) overcome the turbulent energy in plasma heating.}
\label{fig-Kolmogorov-cascade}
\end{figure}

On the other hand, a certain stream of high-energy photons coming from the radiation zone through the tachocline and through the ``ring'' (between the magnetic wall of the flux tube and the O-loop) (see Figs.~\ref{fig-lampochka} and~\ref{fig-Kolmogorov-cascade}) allows to use the barometric equilibrium~\citep{Parker1974a}, i.e. no Parker-Birmann cooling effect, to determine the convective heating $(dQ/dt)_2$ (see Sect.~\ref{sec-radiative-heating}).

\begin{figure*}
\begin{center}
\includegraphics[width=17cm]{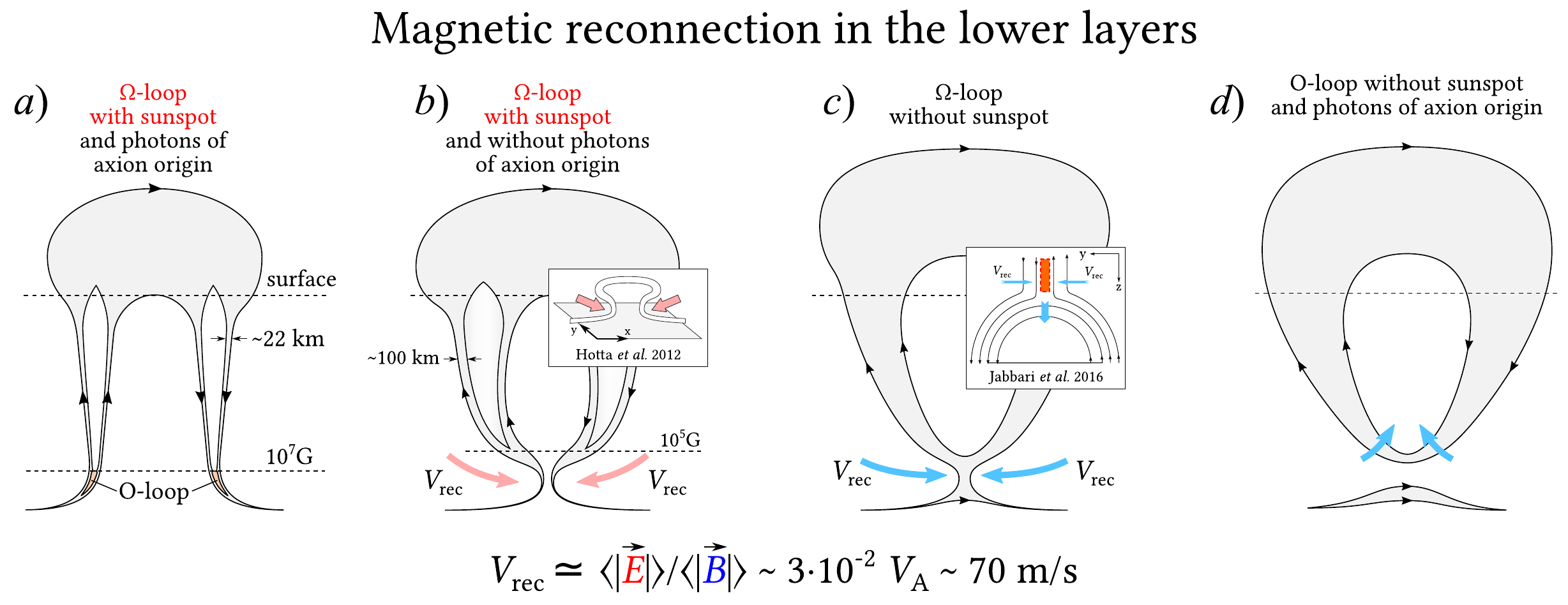}
\end{center}
\caption{A sketch of the magnetic reconnection near the tachocline.
\textbf{(a)} $\Omega$-loop forms the sunspot umbra (with photons of axion
origin from the O-loops) via the indirect thermomagnetic EN~effect but
without reconnection (Fig.~\ref{fig-lampochka});
\textbf{(b)} $\Omega$-loop with a sunspot (without photons of axion origin
and O-loops);
pink arrows show the upward convective flow between the ``legs'' of the
$\Omega$-loop during its rise from the tachocline to the visible surface;
\textbf{(c)} $\Omega$-loop with reconnection and without a sunspot;
\textbf{(d)} O-loop without a sunspot.
Going through the stages (a), (b), (c), (d) (left to right), the convection
around the rising $\Omega$-loop ``closes'' it at its base, then a free O-loop is
formed via reconnection (d), and the initial configuration of the azimuthal field
at the bottom of this region is restored. Blue arrows show the substance motion
leading to the loop ``legs'' connection.}
\label{fig-lower-reconnection}
\end{figure*}

This solution explicitly depends on the lifetime of the magnetic tubes rising
from the tachocline to the solar surface. Therefore, because of the magnetic
reconnection in the lower layers (see Fig.~4 in \cite{Parker1994}), it is not
the final stage of the simulation.
The essence of a virtually empty tube which is first born without a dynamo of any kind (Fig.~\ref{fig-lower-reconnection}a), is related to the physics of the turbulent reconnection of magnetic bipolar structures (see Fig.~\ref{fig-lower-reconnection}b,c and Eq.~(8) in \cite{Jabbari2016}). It is therefore connected to a very rare model of fluctuation dynamo caused by a multiscale turbulence model (see \cite{Baggaley2009}), but necessarily through the so-called turbulent reconnection \citep{Loureiro2009,Huang2010,Beresnyak2017}. In this case the turbulent pumping of the plasma from the azimuthal field is repeated in the $\Omega$-loop again and again (Fig.~\ref{fig-lower-reconnection}).

%\begin{figure*}[tb!]

As described by \cite{Spruit1987,Wilson1990,Parker1994,Parker2009}, the upward
convection flow around the rising $\Omega$-loop brings its ``legs'' together in
such a way that the magnetic field reconnection occurs across this loop. This
cuts off the loop from the azimuthal magnetic field, turning it into an O-loop
(see Fig.~3 in \cite{Spruit1987} and Fig.~4 in \cite{Parker1994}). After that the
azimuthal magnetic field restores its initial configuration and becomes ready
for another process with $\Omega$-loop.

Let us now make some important remarks on the turbulent reconnection, the
$\Omega$-loop transformation into the O-loop by rapid ``legs'' closure, the
restoration of the initial azimuthal field and the preconditions for another
$\Omega$-loop formation in the same place. It is also necessary to explain the
physical interpretation of the overshoot process near the tachocline and
estimate the velocity $v_{rise}$ and time $\tau _{rise}$ of the magnetic tube
rise from the overshoot boundary layer -- starting with the azimuthal
magnetic flux strength of $B_{tacho} \sim 4 \cdot 10^7 ~G$
(see Eq.~A.17 in \cite{RusovDarkUniverse2021}).

One ultimate goal of this section is to determine the general regularities in
the theory of MFTs, which are generated by the magnetic buoyancy
of virtually empty tubes rising from the tachocline to the surface of the Sun
(Fig.~\ref{fig-lampochka}). Another one is
the physical interpretation of the process of MFTs reconnection
in the lower layers of the convection zone (Fig.~\ref{fig-lower-reconnection}).
Not only this is related to the magnetic cycles of flux tubes coinciding with
the observed Joy's law for the tilt angle, but both effects (surprisingly
enough) are induced by the existence of DM -- the solar axions
generated in the core of the Sun.

\subsection{Convective heating and the buoyant rise of magnetic flux tubes: the phenomenon of solar dark matter axions}
\label{sec-radiative-heating}

The first problem is devoted to the study of the effect of virtually empty
magnetic tubes and the phenomenon of solar dark matter axions.

The assumption that the virtually empty magnetic tubes
(Fig.~\ref{fig-lampochka}) are neutrally buoyant ($\rho_{int} = \rho_{ext}$
\citep{Spruit1982}) implies that the temperature inside these tubes is lower than
that of the ambient medium (Fig.~\ref{fig-lampochka} and Fig.~\ref{fig-lower-reconnection}a).
This leads to the heat inflow, and consequently, the flux tube rises up
(see \cite{Parker1975} or Sect.~8.8 in \cite{Parker1979a}). For
a horizontal tube with a cross-section of radius $a$ the rise velocity follows
from the Parker's analysis (see \cite{Parker1975}, Eq.~(60) in 
\cite{Ballegooijen1982}):

\begin{equation}
v_{rise} = 2 \frac{H_p}{\tau _d} \frac{B^2}{8 \pi p_{ext}}
\left( -\delta + 0.12 \frac{B^2}{8 \pi p_{ext}} \right)^{-1},
\label{eq07-39}
\end{equation}

\noindent
where $H_p = \Re T_{ext}/g = p_{ext}/g \rho _{ext} = 0.08 R_{Sun}$ 
\citep{BohmVitense1958,Spruit1977,Brun2011} is the pressure scale height at the
tachocline, $T_{ext}$ and $p_{ext}$ are the external gas temperature and
pressure, $\delta \equiv Y = \nabla _e - \nabla _{ad} = -c_p^{-1} dS / d \xi 
= -c_p^{-1} H_p dS / dz$ is the dimensionless entropy gradient 
(see \cite{Ballegooijen1982,Smolec2008,Smolec2010}),
$\nabla _e \equiv d \ln T_{ext} / d \ln p_{ext}$ and
$\nabla _{ad} \equiv (\partial \ln T / \partial \ln p)_s$ are the local and
adiabatic temperature gradients in external and internal plasma
\citep{Spruit1974,Ballegooijen1982,Christensen1995}, $s$ is the specific
entropy, $c_p$ is the heat capacity at constant pressure, and $\tau _d$ is the
radiation and/or convection diffusion time of the flux tube:

\begin{equation}
\tau_d = \frac{c_p \rho a^2}{k_e} \simeq 
c_p \rho a^2
\left[ \frac{c_p  F_{tot} }{g} 
\left( 1 + \frac{2 \ell _{ov}}{5 H_p} \right)^{\nu} \right]^{-1}.
\label{eq07-40}
\end{equation}

\noindent
where for the fully ionized gas $c_p = 2.5 \Re$ ($\Re$ is the gas constant
in the ideal gas law $p = \rho \Re T$), $T(z)$ and $\rho(z)$ are the mean
temperature and density; $k_e$ is the radiative heat conductivity
(see Eq.~(36) in \cite{Ballegooijen1982});
$\ell _{ov} \approx 0.37 H_p$ \citep{Ballegooijen1982,Christensen2011} is the
thickness of the overshoot layer; the total radiative energy flux $F_{tot} = L/(4 \pi r^2)$ depends on the Sun luminosity $L$; $g$ is the
gravitational acceleration.

Next we apply the condition of hydrostatic equilibrium, $dp / dz = \rho g$,
when the adiabatic temperature gradient $(dT/dz)_{ad} = g/c_p$ may be used,
and the neutral buoyancy of the flux tube in the overshoot zone 
$(\vert \delta T \vert /T_{ext})^{-1} \sim \beta \equiv 8\pi p_{ext}/B^2$. This way we are able to
estimate the time of the radiative and/or convective diffusion $\tau_d$
(see Eq.~(\ref{eq07-40})) of the flux tube:

\begin{equation}
\tau_d = \frac{c_p \rho a^2}{k_e} \approx
\vert \delta T \vert c_p \rho 
\frac{a^2}{(1.148)^{\nu} \delta z F_{tot} }\, ,
\label{eq07-41}
\end{equation}

\noindent where

\begin{equation}
\delta z \sim (1.148)^{-\nu} \left( \frac{a}{H_p} \right)^2
H_p \frac{\nabla _e}{\nabla _{rad}}, ~~ where ~~\nu \geqslant 3.5\, ,
\label{eq07-42}
\end{equation}

\begin{equation}
F_{tot} = \frac{L}{4 \pi R_{tacho}^2} = 
H_p \frac{\nabla _{rad}}{\nabla _e} \left( \frac{dQ}{dt} \right)_1 .
\label{eq07-43}
\end{equation}

\noindent
Here $\nabla _{rad} = (\partial \ln T_{ext} / \partial \ln p_{ext})_{rad}$ is the
radiative equilibrium temperature gradient; $(dQ/dt)_1$ is the rate of radiative heating, which only depends on the
thermodynamic parameters $k_e$ and $T_{ext}$ of the ambient plasma, depending only
on the radial distance from the Sun center \citep{Spruit1974,Ballegooijen1982}.

As a result, it is not difficult to show that the van Ballegooijen model
combining equations (\ref{eq07-39})-(\ref{eq07-43}) gives the final
expression for the rise time by radiation and/or convective diffusion from the
boundary layer of the overshoot to the solar surface,

\begin{equation}
\tau_d \approx \frac{2}{\beta} T_{ext} \left[ \frac{1}{c_p \rho_{ext}} 
\left( \frac{dQ}{dt} \right)_1 \right]^{-1},
\label{eq07-44}
\end{equation}

\noindent
and the lifting speed of the MFT from the overshoot boundary
layer to the surface of the Sun,

\begin{align}
&v_{rise} = H_p \nabla _{ad} \frac{1}{p_{ext}} \left( \frac{dQ}{dt} \right)_1
\left( -\delta + 0.12 \frac{B^2}{8 \pi p_{ext}} \right)^{-1}, \nonumber \\  
&\nabla _{ad} = \nabla _e \simeq 0.4,
\label{eq07-45}
\end{align}

\noindent
which are almost identical to the equations (29) and (30) of \cite{Fan1996}.

Hence, we understand that the van Ballegooijen model is a special case for
magnetic fields of $B_{tacho} \leqslant 10^5~G$, under which a magnetic dynamo
can exist. On the other hand, we know that based on the holographic BL~mechanism, generating (in contrast to dynamo!) the toroidal magnetic
field in the tachocline, the universal model of flux tubes predetermines the
existence of not only the fields of $B_{tacho} \leqslant 10^5~G$, but also
the strong magnetic fields of the order $B_{tacho} \sim 10^7 ~G$.

Unlike the special van Ballegooijen model, we adopt the universal model of MFTs with

\begin{equation}
v_{rise} = 2 \frac{H_p}{\tau _d} \frac{B^2}{8 \pi p_{ext}}
\left( -\delta + 0.12 \frac{B^2}{8 \pi p_{ext}} \right)^{-1}, \nonumber
\end{equation}

\noindent where

\begin{equation}
\tau_d = \frac{c_p \rho a^2}{k_e} 
\left[ 1 + \left( \frac{dQ}{dt} \right)_2 
\middle/ \left( \frac{dQ}{dt} \right)_1 \right]^{-1} .
\label{eq07-46}
\end{equation}

\noindent
Here $(dQ/dt)_2$ represents the radiative diffusion across the
flux tube due to the temperature difference ($\delta T \equiv T - T_{ext}$)
between the tube and the external plasma (see \cite{Fan1996}).

Using simple calculations of equations (\ref{eq07-39}) and (\ref{eq07-46}) for
MFTs, it is easy to show that with the help of the total
expression

\begin{equation}
\frac{dQ}{dt} = \left( \frac{ dQ}{dt} \right)_1 + \left( \frac{ dQ}{dt} \right)_2
\label{eq07-47}
\end{equation}

\noindent
of the universal model

\begin{equation}
\tau_d \approx \frac{2}{\beta} T_{ext} 
\left[ \frac{1}{c_p \rho_{ext}} \frac{dQ}{dt} \right]^{-1}
%\label{eq07-51}
\label{eq07-48}
\end{equation}

\noindent and

\begin{align}
&v_{rise} = H_p \nabla _{ad} \frac{1}{p_{ext}} \frac{dQ}{dt}
\left( -\delta + 0.12 \frac{B^2}{8 \pi p_{ext}} \right)^{-1}, \nonumber \\ 
&\nabla _{ad} = \nabla _e \simeq 0.4,
%\label{eq07-52}
\label{eq07-49}
\end{align}

\noindent
which is the general case of the so-called universal model of van Ballegooijen-Fan-Fisher (vanBFF model).

On the other hand, let us remind that on the basis of the BL
holographic mechanism, generating the toroidal magnetic field in the
tachocline, the universal model of flux tubes is predetermined by the
existence of strong magnetic fields of the order of $B_{tacho} \sim 10^7~G$.
Since the physics of the holographic BL~mechanism does not
involve a magnetic dynamo, we often refer to it as the universal antidynamo
vanBFF model. It is determined by the following total
energy rate per unit volume:

\begin{equation}
\frac{dQ}{dt} = 
\left( \frac{ dQ}{dt} \right)_1 + \left( \frac{ dQ}{dt} \right)_2 = 
\left( \frac{ dQ}{dt} \right)_1 \left[ 1 + \frac{\alpha _1 ^2}{\nabla _e} 
\left( \frac{H_p}{a}  \right)^2 \frac{1}{\beta} \right],
%\label{eq07-53}
\label{eq07-50}
\end{equation}

\noindent where

\begin{equation}
\left( \frac{dQ}{dt} \right)_1 = -\nabla \vec{F}_{rad} = F_{tot}
\frac{\nabla _e}{\nabla _{rad}} \frac{1}{H_p} = k_e \nabla _e
\frac{T_{ext}}{H_p^2} ,
%\label{eq07-54}
\label{eq07-51}
\end{equation}

\begin{equation}
\left( \frac{dQ}{dt} \right) _2 = -k_e \frac{\alpha _1^2}{a^2} (T - T_{ext}) ,
%\label{eq07-55}
\label{eq07-52}
\end{equation}

\noindent
where we used the approximate relation
$\vert \delta T \vert /T_{ext} \sim 1/\beta$); the parameter $\alpha _1 ^2 \approx 5.76$
\citep{Fan1996,Weber2015}; $\vec{F}_{rad}$ is the radiative energy flux 
\citep{Spruit1974}; $\nabla _e \sim 1.287 \nabla _{rad}$
(see Table 2 in \cite{Spruit1974}); $H_p/a$ is a factor for
the lower convection zone (see \cite{Ballegooijen1982}).

As a result, we understand that the heating rate of MFTs
consists of the rate of radiative and convective heating (see 
Eqs.~(\ref{eq07-47})-(\ref{eq07-52}); Eq.~(10) in \cite{Fan1996} and
Eq.~(7) in \cite{Weber2015}):

\begin{equation}
\frac{dQ}{dt} = \rho T \frac{dS}{dt} \approx 
-div \vec{F}_{rad} - k_e \frac{\alpha _1^2}{a^2} (T - T_{ext}) ,
%\label{eq07-57}
\label{eq07-53}
\end{equation}

\noindent
where $S$ is the entropy per unit mass. The first term $(dQ/dt)_1$ determines the mean temperature
gradient between the lower convection zone and the overshoot (see
Eq.~(\ref{eq07-51})). It deviates from the radiative equilibrium substantially,
implying the existence of the nonzero divergence of the heating radiation
flux. The second term $(dQ/dt)_2$ (see Eq.~(\ref{eq07-52})) represents the 
radiation diffusion through the flux tube because of the temperature difference
between the tube and the external plasma. Its effect is to reduce the
temperature difference.

Hence, it is clear that with strong toroidal magnetic fields in the
tachocline (of the order $> 10^5 ~G$) the second term $(dQ/dt)_2$
(in contrast to \cite{Ballegooijen1982,Fan1996,Weber2015,Weber2016})
is the dominant source of convective heating. 
For the toroidal magnetic field 
$B_{tacho} \sim 4 \cdot 10^7 ~G$ in the tachocline, our estimate of
the second term

\begin{align}
\left( \frac{dQ}{dt} \right)_2 = \frac{\alpha _1^2}{\nabla _e} 
\frac{1}{\beta} \left( \frac{H_p}{a} \right)^2 \left( \frac{dQ}{dt} \right)_1
\geqslant &5 \cdot 10^{4} ~erg \cdot cm^{-3} s ^{-1} \nonumber \\
&at ~~ \beta \simeq 1
%\label{eq07-58}
\label{eq07-54}
\end{align}

\noindent
is related to the first term 
$(dQ/dt)_1 \approx 29.7~erg \cdot cm^{-3} \cdot s^{-1}$
(see Eq.~(18) in \cite{Fan1996}) and the following parameters:
$\alpha _1^2 \approx 5.76$ \citep{Fan1996,Weber2015}, 
$\nabla _e \approx 0.4$ (see Table~2 in \cite{Spruit1974,Fan1996}),
$1 / \beta = B_{tacho}^2 / 8 \pi p_{ext} = 1$,
$\nabla _e / \nabla _{rad} \approx 1.287$, 
$a = (\Phi / \pi B_{tacho})^{1/2} \leqslant 0.1 H_p$ (see \cite{Fan1993})
with an average value of magnetic flux $\Phi \sim 10^{21}~Mx$ (see e.g. 
\cite{Zwaan1987}).

Taking into account the consequences of the non-local theory of mixing
length \citep{Spruit1974,Spruit1982}, it can be shown that thin,
neutrally buoyant flux tubes are stable in the stably stratified medium,
provided that its field strength $B$ is smaller than a critical value $B_c$
\citep{Ballegooijen1982}, which is approximately given by

\begin{equation}
\frac{B _c^2}{8\pi p_{ext}} = - \gamma \delta = - \frac{5}{3} \delta .
%\label{eq07-60}
\label{eq07-55}
\end{equation}

So, for the maximum value of the toroidal magnetic field
($B_{tacho} \sim 4 \cdot 10^7 ~G$) the estimate of the rise time of the
radiation and/or convection diffusion of the flux tube (see 
Eq.~(\ref{eq07-48}))

\begin{equation}
\tau _d \approx \frac{2}{\beta} T_{ext} 
\left[ \frac{1}{c_p \rho _{ext}} \left( \frac{dQ}{dt} \right)_2\,\right] ^{-1}
\geqslant 10^4 ~year
%\label{eq07-63}
\label{eq07-56}
\end{equation}

\noindent
and the lifting speed of an almost empty magnetic tube (see Eqs.~(\ref{eq07-54}), (\ref{eq07-55}))

\begin{equation}
v_{rise} \simeq \frac{H_p \nabla _e}{p_{ext}}  \left( \frac{dQ}{dt} \right)_2
\frac{1}{0.72} \leqslant 10^{-8} ~km/s
%\label{eq07-61}
\label{eq07-57}
\end{equation}

\noindent
show that the existence of MFTs on the surface of the Sun is
meaningless.

In this regard,
there is one more beautiful task associated with our problem of
almost total suppression of radiative heating in virtually empty magnetic
tubes (see Fig.~\ref{fig-lampochka}). Let us remind that photons going from the
radiation zone through the horizontal field of the O-loop near the tachocline
(Fig.~\ref{fig-lampochka}) are turned into
axions, thus almost completely eliminating the radiative heating in the
virtually empty magnetic tube. Some small photon flux can still pass
through the ``ring'' between the O-loop and the tube walls (see
Fig.~\ref{fig-lampochka}) and reach the penumbra.
Let us denote the area of the magnetic tube ``ring'' by $2\pi r a _{conv}$,
where $r$ is the magnetic tube radius, and $a_{conv}$ is the width of the
``ring''.

%\begin{figure}[tpb!]
\begin{figure}
\centering
\includegraphics[width=8cm]{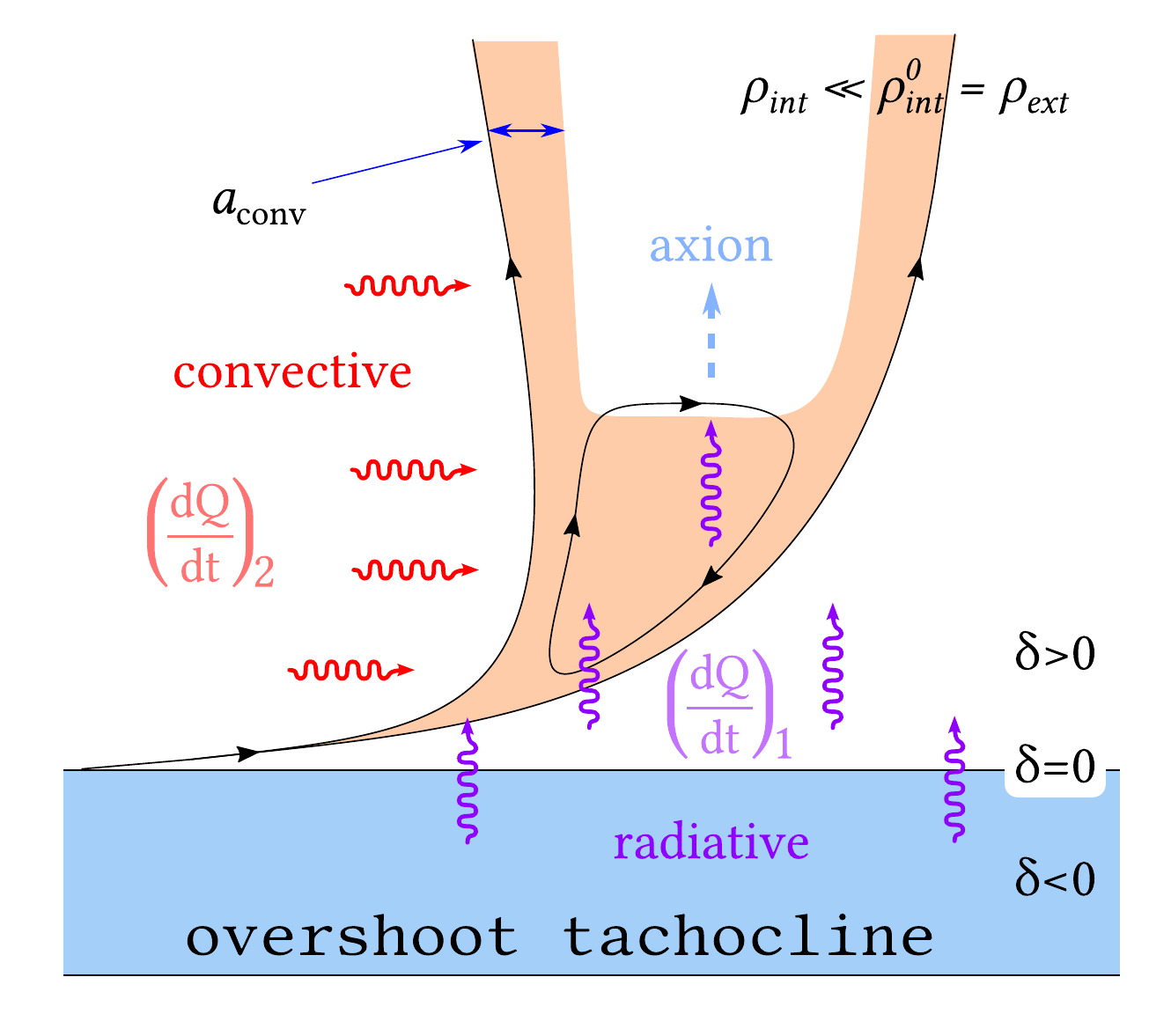}
\caption{A sketch of the magnetic tube born anchored to the
tachocline and risen to the solar surface by the neutral buoyancy
($\rho _{ext} = \rho _{int}^0$). The strong convection suppression inside the
tube leads to the abrupt decrease of temperature and density
($\rho_{int}^0 \gg \rho_{int}$), which in its turn leads to the significant
decrease in gas pressure in the umbra. At the top of the overshoot tachocline
the first term $(dQ/dt)_1$ characterizes the radiative heating which depends on
the thermodynamic quantities $k_e$ and $T_{ext}$ of the external plasma,
changing with the distance from the center of the Sun only (see 
Eq.~(\ref{eq07-51})). The second term $(dQ/dt)_2$ represents the diffuse
radiation (by convective heating) through the flux tube because of the
temperature difference between the tube and the ambient plasma (see 
Eq.~(\ref{eq07-52})). The keV photons (see Fig.~2 in \cite{Bailey2009}) coming
from the radiation zone are turned into axions in the horizontal magnetic field
of the O-loop (see Fig.~\ref{fig-lampochka}). Therefore, the radiative heating
almost vanishes in the virtually empty magnetic tube. The base of the
convection zone is defined as a radius at which the stratification switches
from almost adiabatic ($\delta = \nabla _e - \nabla _{ad} = 0$) to
sub-adiabatic ($\delta = \nabla _e - \nabla_{ad} < 0$). Meanwhile, the external
plasma turns from sub-adiabatic to super-adiabatic
($\delta = \nabla_e - \nabla _{ad} > 0$).}
\label{fig-lower-heating}
\end{figure}

What is the physics behind the appearance of the ``ring'' between the O-loop
and the walls of the magnetic tube (Fig.~\ref{fig-lower-heating})? In simple
words, one can say the following. The appearance of the ``ring'' cross-section
is the result of the production of both axion-origin photons (by converting
solar axions into photons in the tachocline) and photon-origin axions (through
the conversion of high-energy photons from the radiation zone to axions in the
tachocline). From here, on the one hand, the axions of photonic origin are the
part of the manifestation of the mean free path of axion origin photons from
the tachocline to the photosphere. On the other hand, they are the part of the
manifestation of magnetic tube ``ring'', where the convective heating
$(dQ/dt)_2$ dominates over the radiative heating $(dQ/dt)_1$.

As a result, assuming the mean width of the ``ring''
\begin{equation}
a_{conv} \sim 3.7 \cdot 10^{-4} ~H_p,
%\label{eq07-68}
\label{eq07-58}
\end{equation}

\noindent
we apply a new analysis of the universal vanBFF model
(see Eq.~(\ref{eq07-49}) and (\ref{eq07-47}) (analogous to 
\cite{Ballegooijen1982}) or (\ref{eq07-49}) and (\ref{eq07-55}) (analogous to
Eq.~(29) in \cite{Fan1996})), where the calculated values such as the magnetic
flux $\Phi$ and the rise speed $(v_{rise})_{conv}$ of the MFT
to the surface of the Sun, do not contradict the known observational data:

\begin{itemize}

\item the value of the magnetic flux of the tube (for $r \sim 70~km$)
\begin{equation}
\Phi = 2\pi r a_{conv} B_{tacho} \approx 3.7 \cdot 10^{21} Mx ,
%\label{eq07-69}
\label{eq07-59}
\end{equation}

\noindent
which is in good agreement with the observational data of Zwaan~\cite{Zwaan1987};

\item the time of radiation diffusion of the flux tube (\ref{eq07-48})-(\ref{eq07-50})

\begin{align}
(\tau_d)_{conv} &= 2T_{ext} \left \lbrace \frac{1}{c_p \rho_{ext}} 
\left( \frac{dQ}{dt} \right)_1
\left[ 1 + \frac{\alpha _1 ^2}{\nabla_e} 
\left( \frac{H_p}{a_{conv}} \right)^2 
\right] \right \rbrace ^{-1}     \nonumber \\
& \approx 1.1 \cdot 10^5 ~s ~ \sim 1.3 ~ day
%\label{eq07-70}
\label{eq07-60}
\end{align}

\noindent
and, as a consequence, the magnitude of the lifting speed of the MFT (see Eq.~(\ref{eq07-39}))

\begin{equation}
(v_{rise})_{conv} = \frac{2.77 H_p}{(\tau_d)_{conv}} \sim 1.4 ~km/s,
%\label{eq07-71}
\label{eq07-61}
\end{equation}

\noindent
which are almost identical to the observational data of the known works by
%\cite{Ilonidis2011,Ilonidis2012,Ilonidis2013} and 
Kosovichev~\textit{et~al.}~\cite{Kosovichev2016}.
\end{itemize}

In order to show signatures of appearing areas of sunspots inside the Sun before they appear on the surface, we used, according to Kosovichev~\textit{et~al.}~\cite{Kosovichev2016}, the measurements of plasma flows in the upper convection zone, provided by the Time-Distance Helioseismology pipeline developed for the analysis of solar oscillation data obtained by the Helioseismic and Magnetic Imager (HMI) within the Solar Dynamics Observatory (SDO), to investigate the subsurface dynamics of the emerging active region (AR) NOAA 11726.

\begin{figure}
\centering
\includegraphics[width=14cm]{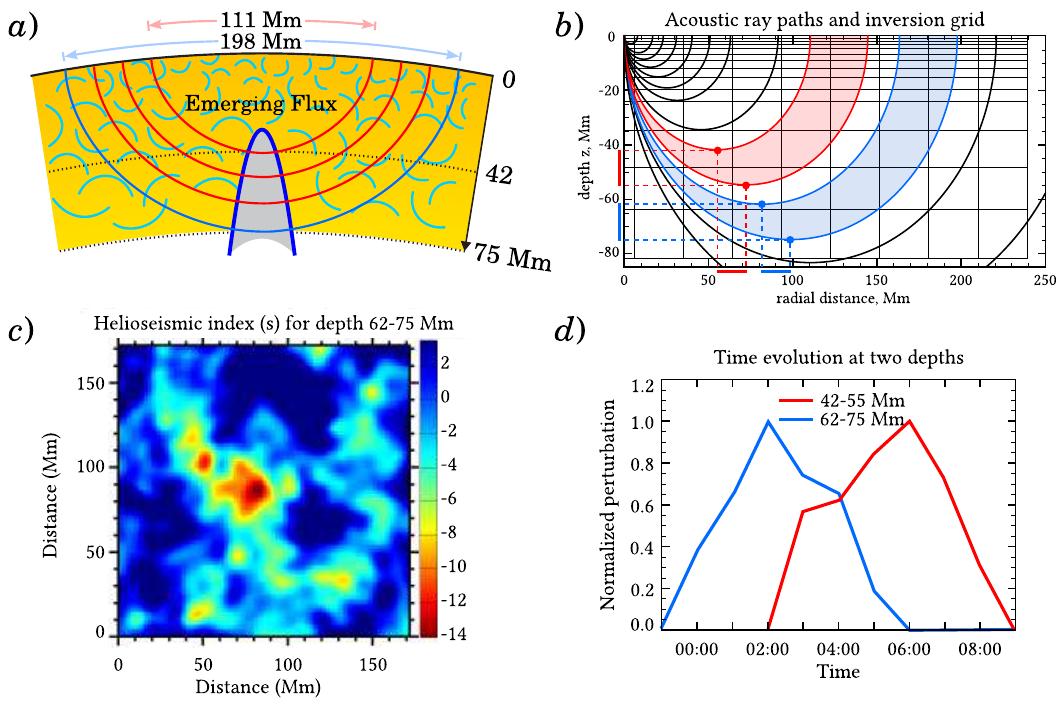}
\caption{Detection of emerging active region AR 11726: 
\textbf{a)} magnetograms of helioseismic and magnetic imager (HMI) on board of the Solar Dynamics Observatory (SDO) of April 19, 2013, 3:00~UT, prior the emergence; the rings show the area where the acoustic oscillation signal was measured to detect the subsurface signal at the central point;
\textbf{b)} a vertical cut through the computational grid used in the time-distance inversions, and a sample of ray paths of acoustic waves, which were used for measuring travel times~\cite{Kosovichev2008};
\textbf{c)} variations of a travel-time index showing the AR perturbation located the depth of 62-75~Mm on April~19, 2013, 3:00~UT;
\textbf{d)} variations of the helioseismic perturbation associated with the emerging AR at two different depths: 42-55~Mm and 62-75~Mm, as a function of time. The characteristic rise time estimated using the delays in two layers at depths 42-55~Mm and 62-75~Mm is about 1.4~km/s. Derived from~\cite{Kosovichev2016}.}
\label{fig-emerging-flux}
\end{figure}

Figure~\ref{fig-emerging-flux}c shows the distribution of the effective phase shift (``helioseismic index'') for the depth range 62-75~Mm measured on April 19, 2013, 03:00~UT, when there was no significant magnetic flux on the surface. The figure~\ref{fig-emerging-flux}d shows the variations of the helioseismic perturbation associated with the emerging AR at two different depths: 62-75~Mm and 42-55~Mm depending on time. One of the greatest advantages of the remote helioseismology is that the full-disk observations of solar variations allow to build a special scheme of measurement and selection of acoustic waves signals (Fig.~\ref{fig-emerging-flux}b), passing though certain sub-surface areas. For example, Ilonidis~\textit{et~al.}~\cite{Ilonidis2013} developed a special proceadure of deep focusing which is able to detect large emerging active areas more than a day before their appearance on the surface (Fig.~\ref{fig-emerging-flux}). The rise velocity determined by tracking the magnetic flux declination with depth, is about 1.4~km/s, which is very close to the rise velocity in the deep layers~\cite{Kosovichev2016}, e.g. near tachocline!
Interestingly, the results of several attempts to detect the emerging magnetic flux before its appearance in the photosphere may be compared to our theoretical estimates (\ref{eq07-60})-(\ref{eq07-61}) and demonstrate a surprising agreement in terms of the magnetic tube rising velocity and time.

At the same time, it is known that observations of the characteristic rising speed of active regions (sunspot groups) $\sim 1.0-1.4~km/s$~\cite{Ilonidis2013,Kosovichev2000,Zharkov2008,Toriumi2013,Kosovichev2016}, which coincides with a speed of $\sim 1.4~km/s$ of the universal model of van~Ballegooijen-Fan-Fisher (see vanBFF model for (\ref{eq07-49})), is an order of magnitude larger than the speed of 0.1-0.25~km/s, associated with numerical models of the emerging flux (see~\cite{Fan2021}).

Below we briefly and simply show how our fundamental physics of the rise of magnetic field tubes based on solar axions of dark matter differs from the classical physics of an active region emerging flux from convective dynamos in the bulk of the solar convection zone (see e.g.~\cite{Fan2021,Charbonneau2020,Hotta2020}).

Let us first draw some attention to what the remarkable physicist Yuhong~Fan says about the essence of magnetic fields in the solar convection zone: ``\textit{The question of whether, or to what extent, a strong toroidal magnetic field stored in the overshoot region at the base of the convection zone, generated by a deep seated solar dynamo process, is responsible for the formation of solar active regions remains to be more rigorously investigated.}''~\cite{Fan2021}

In contrast to the classical physics of an active region emerging flux from convective dynamos in the bulk of the solar convection zone (see e.g.~\cite{Fan2021,Charbonneau2020,Hotta2020}), we already know that:

\begin{itemize}
\item first, a strong toroidal magnetic field, stored in the overshoot tachocline region, is predetermined by the thermomagnetic Ettingshausen-Nernst effect (see Apendix~A in~\cite{RusovDarkUniverse2021}), which exactly (in contrast to dynamo!) coincides with $B_{tacho} = 4\cdot 10^7~G$; at the same time, we indirectly showed that, using the holographic principle of quantum gravity (see Apendix~C in~\cite{RusovDarkUniverse2021}), the repelling toroidal magnetic field of the tachocline exactly ``neutralizes'' the magnetic field in the Sun core~\cite{Fowler1955,Couvidat2003}, where the projections of the magnetic fields of the tachocline and the core have the equal value but the opposite directions.

\item second, unlike the component of the solar dynamo model (see Fig.~C.1a in~\cite{RusovDarkUniverse2021} and \cite{Fan2021,Charbonneau2020}), the Babcock-Leighton mechanism (see Fig.~C.1b in~\cite{RusovDarkUniverse2021}), predefined by the the fundamental holographic principle of quantum gravity, and thermomagnetic EN effect, emphasizes that this process is associated with continuous transformation of toroidal magnetic energy into poloidal magnetic energy ($T \rightarrow P$ transformation), but not vice versa ($P \rightarrow T$). It means that the holographic BL mechanism is the main process of regeneration of the primary toroidal field in the tachocline, and thus, the formation of buoyant toroidal magnetic flux tubes at the base of the convective zone, which then rise to the surface of the Sun.

\item third, our physics of the magnetic flux tubes rise is related to the fact that if the axions, which are born in the Sun's core, are directly converted into X-rays near the tachocline, then the axion-photon oscillations predetermine the appearance of magnetic sunspot cycles. This is due to the fact that the formation of sunspots and their cycles is a consequence, according to~\cite{RusovDarkUniverse2021}, of the anticorrelation 11-year cycles of ADM density modulation inside the Sun, which, not surprisingly, are a consequence of the 11-year ADM halo density modulation in the fundamental plane of the galactic center, which closely correlates with the density modulation of baryonic matter near the supermassive black hole. From here, it is easy to show (see~\cite{RusovDarkUniverse2021}) in what way the anticorrelation identity between indicators of the ADM density modulation inside the Sun and the number of sunspots (or the correlation identity between indicators of modulation of solar axions (or photons of axion origin) and sunspot cycles) is realized.
\end{itemize}

Finally, let us show that if the rise velocity of the floating magnetic tube, determined by the magnetic field values $\sim 4.1\cdot 10^7~G$ and the ``ring'' width $\sim 3.7\cdot 10^{-4} ~H_p$, causes the appearance of MHT in the form of sunspots, then the parameters of the universal vanBFF model will be associated with the magnetic cycles that are almost identical to the observational data of the tilt angle of Joy's law.

\subsection{Secondary reconnection of magnetic tubes in lower layers and the observed features of the tilt angle of Joy's Law}
\label{sec-tilt}

The problem is devoted to physics of the magnetic reconnection of a
magnetic tube in the lower layers, which is associated with the so-called
reconnecting dynamo and the observed features of the tilt angle of Joy's law.

In contrast to the primary reconnection inside magnetic tubes near the
tachocline, which generates the O-loops, thus participating in the formation of
photons of axion origin, as well as the axions of photon origin, from the
O-loop to the photosphere, here we are interested in the secondary magnetic
reconnection, which, although does not depend on sunspot cycles, but is
necessarily related to the process of sunspots liftoff from the surface of the
Sun.

\begin{figure*}
\begin{center}
\includegraphics[width=14cm]{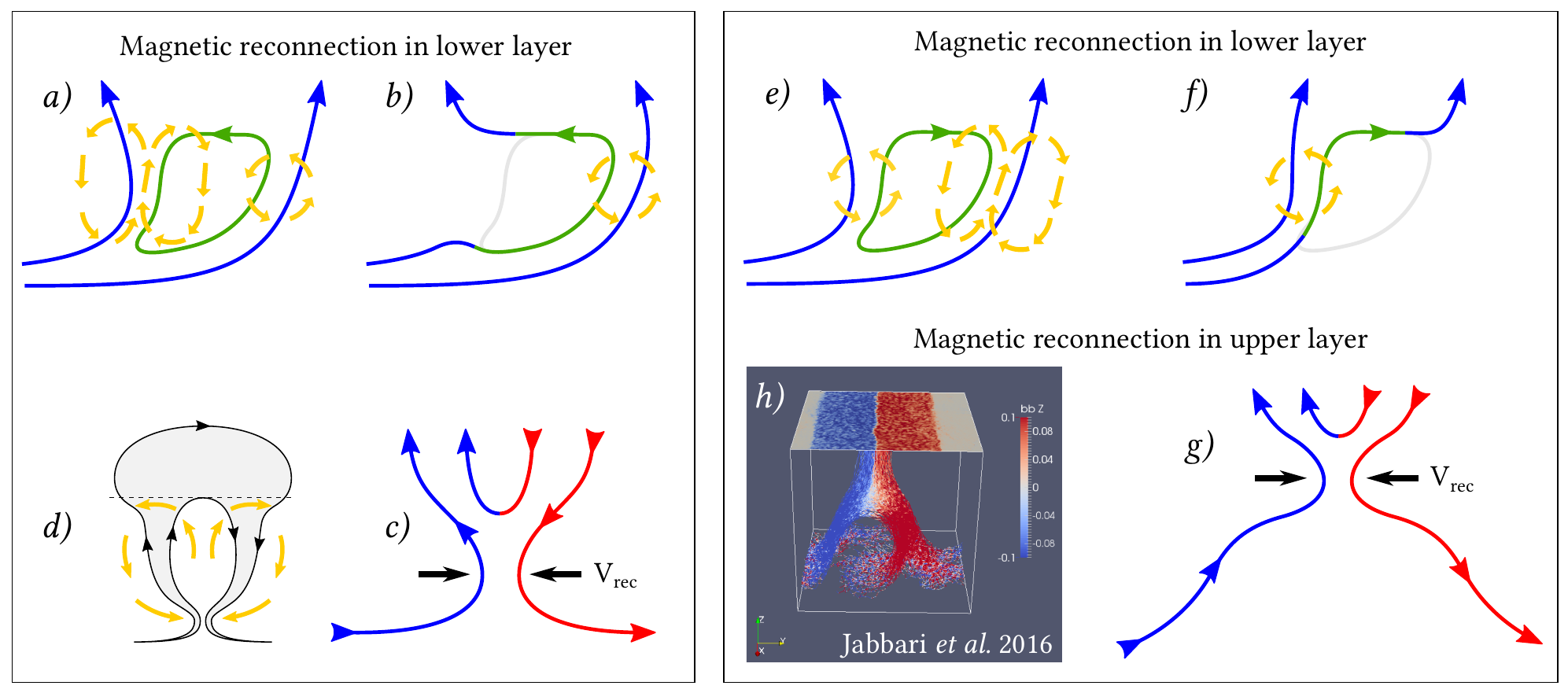}
\end{center}
\caption{Topological effects of magnetic reconnection in the lower (left) or upper (right) layers of the magnetic tube. Here, the unipolar part of the $\Omega$-loop is rebuilt on its base, compressing the $\Omega$-loop (blue lines) to form a free O-loop (green lines) (Fig.~\ref{fig-lower-reconnection}a). The yellow lines show the movement of the substance leading to the connection of the ``leg'' loop (see analogous Fig.~4 in \cite{Parker1994}). If the O-loop (green lines) can randomly have different directions of magnetic fields, then the magnetic reconnection can generate loop ``legs'' in different layers, for example, in lower layers (Fig.~\ref{fig-upper-reconnection}c and also Fig.~\ref{fig-upper-reconnection}d as an analog of Fig.~\ref{fig-lower-reconnection}b) and upper layers (Fig.~\ref{fig-upper-reconnection}g and also Fig.~\ref{fig-upper-reconnection}h as an analog of Fig.~4 in \cite{JabbariEtAl2016}).}
\label{fig-upper-reconnection}
\end{figure*}

In Fig.~\ref{fig-upper-reconnection} one can see the conditions for the secondary reconnection between the O-loop (green lines) and the unipolar part of the $\Omega$-loop (blue lines) that can organize them in the lower layer (Fig.~\ref{fig-upper-reconnection}, left) or in the upper layer (Fig.~\ref{fig-upper-reconnection}, right), thereby showing the appearance of bipolar magnetic tubes in various versions (Fig.~\ref{fig-upper-reconnection}c,d and g,h).

%\begin{figure*}[tbp]

\begin{figure*}
\begin{center}
\includegraphics[width=17cm]{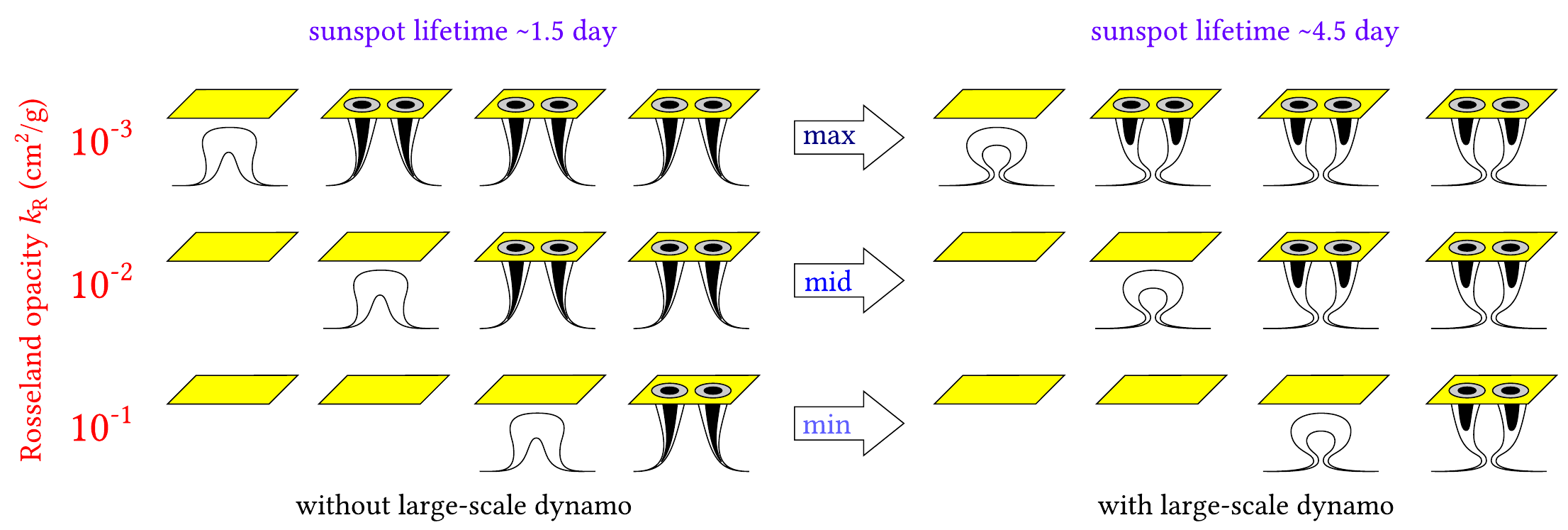}
\end{center}
\caption{The physical nature of the cycle of sunspots as a consequence of the modulation of MFTs, rising from the tachocline to the surface of the Sun. Left: ``one'' of the initial bipolar magnetic tubes is ``invisible'' due to the absence of a free O-loop and, thus, the absence of photons of axion origin in the $\Omega$-loop, which, as a consequence, rises super-slowly to the solar surface: here the rise time of the magnetic tube from the boundary layer of the overshoot to the surface of the Sun is many orders of magnitude greater than the lifetime of the sunspot (see \cite{Petrovay1997,Solanki2003} and vanBFF model (Sect.~\ref{sec-radiative-heating})). Right: simplifying the drawing, we drew different unipolar magnetic tubes that actually correspond to visible or ``invisible'' bipolar magnetic tubes, cyclically (from the maximum to the minimum of activity) appearing on the solar surface. The rough values of the Rosseland opacity $k_R$ can be estimated using Eq.~B.4 and Fig.~B.1 in~\cite{RusovDarkUniverse2021}.}
\label{fig-sunspots-vanish}
\end{figure*}

Finally, let us note that we have two types of primary magnetic reconnection inside the $\Omega$-loop near the tachocline, one of which is manifested by the existence of a free O-loop (red lines in Fig.~\ref{fig-Kolmogorov-cascade}, green lines in Figs.~\ref{fig-upper-reconnection}a and ~\ref{fig-upper-reconnection}e), and the other one practically does not exist. In other words, magnetic reconnection inside the $\Omega$-loop may or may not accidentally give birth to a free O-loop near the tachocline (see Fig.~\ref{fig-Kolmogorov-cascade}).
This means that if a free O-loop is created in the $\Omega$-loop near the tachocline via the primary reconnection, classical sunspots appear on the surface (see black tubes, Fig.~\ref{fig-sunspots-vanish}), while the rare ``transparent'' bipolar tubes, not reaching the surface and having no O-loop inside, appear as ``optically invisible'' spots on the surface (see white bipolar tubes, Fig.~\ref{fig-sunspots-vanish}).

%\begin{figure*}[tbp]

Here arises a question: How does the secondary magnetic reconnection in the lower layers of flux tubes (with magnetic field strength $\sim 10^7 ~G$ (see Fig.~\ref{fig-lower-reconnection}a) and $\sim 10^5 ~G$ (see Fig.~\ref{fig-lower-reconnection}b), respectively) explain the physics and theoretical estimates of the buoyant tubes -- the time and speed of the sunspot liftoff from the solar surface, and the tendency of the tilt angle of Joy's law?

\begin{figure}
\begin{center}
\includegraphics[width=12cm]{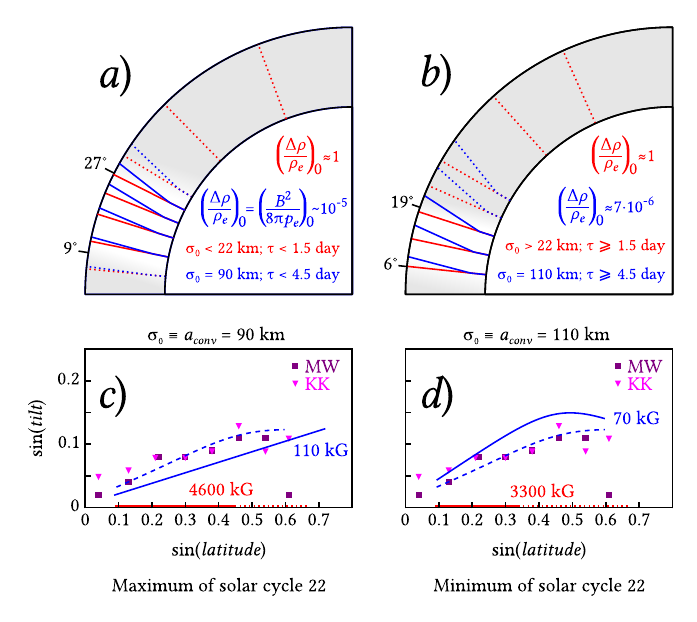}
\end{center}
%\vskip-1cm
\caption{\textbf{(a)}-\textbf{(b)} Flux tubes without drag of adiabatic flux ring in the superadiabatic convection zone (red lines: based on the \cite{Choudhuri1987} and \cite{Choudhuri1989} MEQ model (see also (10) in \cite{DSilvaChoudhuri1993}), as well as on the vanBFF MEQ model (see Sect.~\ref{sec-radiative-heating})), and the trajectories of flux rings in thermal equilibrium (TEQ) incorporating drag (blue lines: based on the TEQ model of \cite{Choudhuri1987} and \cite{Choudhuri1989}; see also (5)-(10) in \cite{DSilvaChoudhuri1993}). In (a) we use the values for the maximum of the magnetic cycle (red lines) of $(\Delta \rho / \rho_{ext})_0 = B_0^2 / 8 \pi p_{ext,0} \approx 1$ and $B_0 \approx 4.6 \cdot 10^7 ~G$ for the $\Omega$-loop with a sunspot and photons of axion origin (see Figs.~\ref{fig-lower-reconnection}a and~\ref{fig-upper-reconnection}a) and (blue lines) of $(\Delta \rho / \rho_{ext})_0 \approx 10^{-5}$ and $B_0 \approx 1.1 \cdot 10^7 ~G$ for the $\Omega$-loop with a sunspot and without photons of axion origin (see Fig.~\ref{fig-upper-reconnection}b). In (b) the values correspond to the minimum of the magnetic cycle: (red lines)  $(\Delta \rho / \rho_{ext})_0 \approx 1$ and $B_0 \approx 3.3 \cdot 10^7 ~G$ for the $\Omega$-loop with a sunspot and photons of axion origin (see Fig.~\ref{fig-upper-reconnection}a) and (blue lines) $(\Delta \rho / \rho_{ext})_0 \approx 2 \cdot 10^{-6}$ and $B_0 \approx 10^7 ~G$ for the $\Omega$-loop with a sunspot and without photons of axion origin (see Fig.~\ref{fig-lower-reconnection}b).
In each of the two cases, stream rings are calculated at latitudes 5$^\circ$, 10$^\circ$, 20$^\circ$, 30$^\circ$, 45$^\circ$, 60$^\circ$.
In (a) and (b) we use red and blue dotted lines of trajectories that depend on large (see Eqs.~(17)-(19) in \cite{Choudhuri1987}, where the diffusion coefficient depends on the density and $k_R$) and small values of the Rosseland mean opacity $k_R$ (see (B.1) and Fig.~B.3 in~\cite{RusovDarkUniverse2021}). 
\textbf{(c)}-\textbf{(d)} Dependence of $\sin (tilt)$ on $\sin (latitude)$ in different theoretical and experimental data series. At an average value of $4.1 \cdot 10^7 ~G$ in the tachocline, the maximum ($\sim 4.6 \cdot 10^7 ~G$) and minimum ($\sim 3.6 \cdot 10^7 ~G$) of the magnetic field of flux tubes are predetermined by two bound estimates, for example, the observational data on the variations of the magnetic field of tubes on the solar surface 
(see \cite{Pevtsov2011,Pevtsov2014}) and the theoretical estimates of magnetic variations in the tachocline with the EN~effect (see Eqs.~A.9 and~A.15 in \cite{RusovDarkUniverse2021}). At an average value of $\sim 10^5 ~G$ near the tachocline, the maximum ($\sim 1.1 \cdot 10^5 ~G$) and minimum ($\sim 7.0 \cdot 10^4 ~G$) of the magnetic field of flux tubes are predetermined by two analogous bound estimates. The dashed lines show the linear regressions of the average slopes of sunspot groups in the latitude range of five degrees for Mount Wilson (MW) and Kodaikanal (KK) observatories. The solid blue lines show theoretical calculations of Joy's law (see \cite{Ivanov2012} and more explanation in the text).}
\label{fig-magtube-tilt}
\end{figure}

Let us first note the physics of primary reconnection in the lower layers of the magnetic flux tubes (with  $B \sim 10^7 ~G$).
Based on the mechanical equilibrium model (MEQ) of \cite{Choudhuri1987} and vanBFF MEQ model (see Sect.~\ref{sec-radiative-heating}), we showed that the results of the simulation of flux tube trajectories without adiabatic ring flux drag in the super-adiabatic zone (see Fig.~\ref{fig-lower-heating}) are well represented in Fig.~\ref{fig-magtube-tilt}a,b (red lines). The most intriguing results of the simulation of flux tube trajectories are a very strong basis not only for understanding the complex physics, but also for understanding the theoretical estimates of buoyant MFTs, like the rise time and speed of rising to the Sun surface (see (\ref{eq07-60})-(\ref{eq07-61}), and also analogous Fig.~1 and Fig.~5 in \cite{Browning2016}), and the explanation of the tendency of the tilt angle of Joy's law (see Fig.~\ref{fig-magtube-tilt}c,d), which do not contradict the known experimental data, for example, the well-known works of
\cite{DasiEspuig2010,DasiEspuig2013,Ivanov2012,McClintock2013,Pevtsov2014,Tlatova2015,Pavai2015,Baranyi2015,Wang2015,Wang2017}.

%\begin{figure}[tbp!]

Further, 
we are interested in the second stage of modeling the trajectory of magnetic flux tubes (see Fig.~\ref{fig-magtube-tilt}), when a magnetic tube that reaches the solar surface above the photosphere changes its shape and structure with the topological effect of the secondary magnetic reconnection in the lower layers of the magnetic tube, see Fig.~\ref{fig-lower-reconnection}b and~\ref{fig-upper-reconnection}a-c. Moreover, we believe that the secondary magnetic reconnection of the flux tubes leads to the real decrease in the magnetic field to $B \sim 10^5 ~G$ at $\sim 0.8~R_{Sun}$ (see Fig.~\ref{fig-Bz-diffusivity}), at which the evolution of the tubes (in thermal equilibrium (TEQ) with the surroundings it is often referred to as magnetic buoyancy \citep{Parker1975}) is controlled by the latitudinal pressure gradient in magnetic layers on the overshoot tachocline that allows a balance between nonzero buoyancy force, curvature force and pressure force in the absence of azimuthal flow (see bottom panel in Fig.~2 in \cite{Schussler2002}; see also review and Fig.~5b in \cite{Fan2009}), which generally allows a balance of TEQ between four main forces: nonzero buoyancy, magnetic tension, aerodynamic drag, and Coriolis force.

\begin{figure}
\begin{center}
\includegraphics[width=12cm]{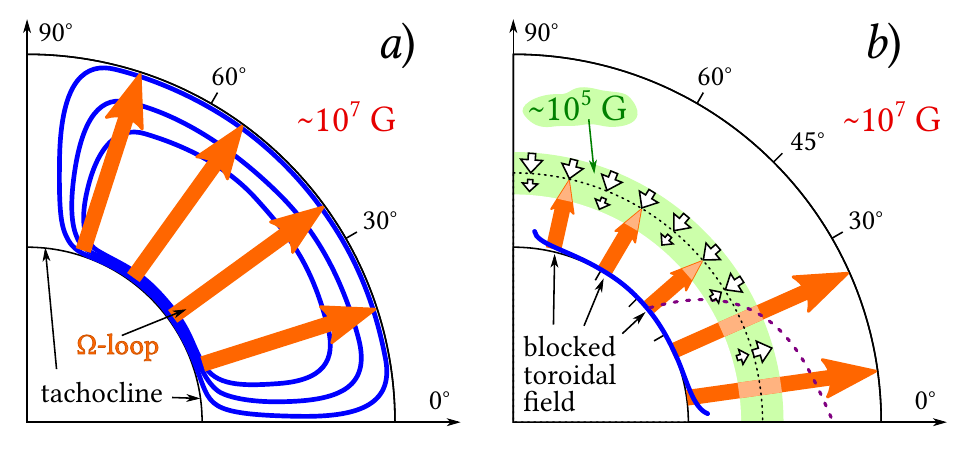}
\end{center}
\caption{Scheme of turbulent reconstruction of the toroidal magnetic field in the convective zone: \textbf{(a)} meridional circulation (closed blue poloidal field lines), which is generated by the toroidal field in the tachocline (thin black lines), and the magnetic buoyancy of the flux tubes (red arrows) with magnetic field $\sim 10^7 ~G$;
\textbf{(b)} the joint interactions of magnetic buoyancy (red arrows) and the rotating magnetic $\nabla \rho$-pumping (short white arrows) by means of reconnection in $\Omega$-loop (Fig.~\ref{fig-Bz-diffusivity}) generate the total buoyancy of magnetic tubes in which the toroidal magnetic field $\sim 10^5 ~G$ predetermines the appearance of magnetic buoyancy on the surface of the Sun only at low, and to a lesser extent at middle latitudes (see analogous Fig.~5 in \cite{Krivodubskij2005}). The dashed curve corresponds to the zero velocity line $v_{dens}^{light} (z, \theta)=0$, where the direction of the magnetic $\nabla \rho$-pumping occurs.}
\label{fig-meridional-cut}
\end{figure}

\begin{figure*}
\begin{center}
\includegraphics[width=14cm]{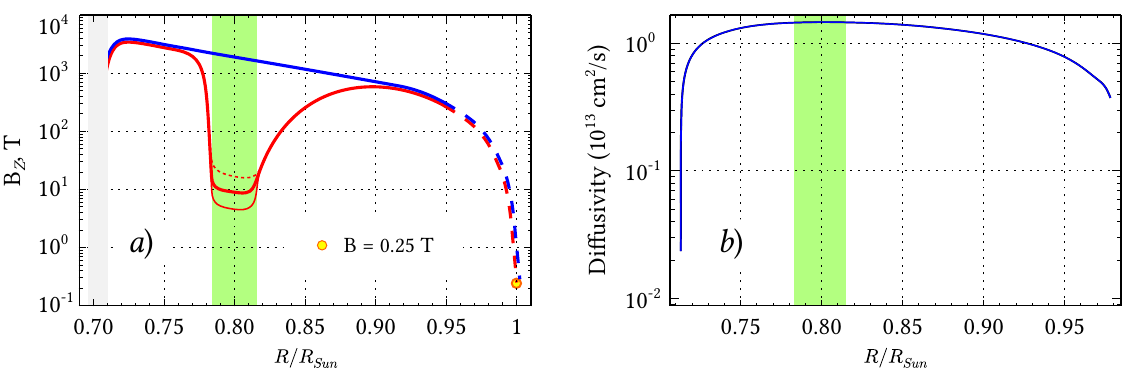}
\end{center}
\caption{\textbf{(a)} Change in the magnetic field strength $B_z$ along the rising $\Omega$-loop as a function of the Sun depth $R/R_{Sun}$ in the convective zone. The blue line (see also Fig.~8 in \cite{Rusov2015,RusovArxiv2019}) denotes admissible values for the standard solar model with diffusion of helium \citep{Bahcall1992}, with the initial value of the theoretical estimate of the magnetic field $B_z \approx  4 \cdot 10^7 ~G$ (see Eq.~(\ref{eq05-002})). The red line corresponds to the cool areas inside the magnetic tube, which by means of the reconnection of the $\Omega$-loop keeps the emerging magnetic diffusivity (green band). \textbf{(b)} Radial profile of turbulent magnetic diffusivity in the solar convection zone based on the \cite{Stix1990} model (see also~\cite{Stix2004}). Diamagnetic pumping must be very strong near the base of the convective zone, where diffusivity almost jumps by orders of magnitude. Its gradient is the rate of descending diamagnetic pumping (see the beginning of the pumping (green band)).}
\label{fig-Bz-diffusivity}
\end{figure*}

Despite our complicated theoretical calculations of the trajectories (Fig.~\ref{fig-magtube-tilt}a,b) of the flux tubes without drag of adiabatic flux ring in the superadiabatic convection zone (red lines: based on the MEQ model by \cite{Choudhuri1987} and the vanBFF MEQ model (see Sect.~\ref{sec-radiative-heating})) and involving the flux ring resistance in thermal equilibrium (blue lines: based on the TEQ model by \cite{Choudhuri1987}), below we will describe the scenario, which explains the simple physics of why the magnetic tubes emerging on the solar surface, can only be at low, and to a lesser extent at middle latitudes (see Figs.~\ref{fig-magtube-tilt}a,b and~\ref{fig-meridional-cut}b).

%\begin{figure}[tbp]

The first part of the scenario consists in discussing the physics of magnetic flux buoyancy and, as a consequence, estimating the rise speed of an almost empty magnetic tube, $v_{rise}$, based on the vanBFF model (see Sect.~\ref{sec-radiative-heating}), which simultaneously coincides with the known expressions for the buoyancy speed, $v_{B}$ (see e.g. \cite{Parker1975}; see also Eq.~(17) and Fig.~2 in \cite{Ballegooijen1988}; Eq.~(34) in \cite{Fan1993}; Eq.~(30) in \cite{Khaibrakhmanov2017}):

\begin{equation}
(v_{rise})_{conv} \equiv (v_{B})_{conv} \approx v_A 
\left( \frac{\pi}{C_D} \frac{a_{conv}}{H_p} \frac{z}{H_p} \right) ^{1/2} ,
\label{eq07-63}
\end{equation}

\noindent where $v_A \equiv B / (4 \pi \rho_{int})^{1/2}$ is the Alfv\'{e}n speed of the magnetic field (see e.g. \cite{Roberts2001}), $C_D \approx 1$ is the drag coefficient, $\rho_{int}$ is the internal density of the gas in the MFT. It is assumed that the magnetic tube is formed inside the disk at an altitude $z / H_p \approx 1$.

For the neutral buoyancy condition ($\rho_{int} = \rho_{ext} \approx 0.2 ~g/cm^3$; see Figs.~6 and 11 in \cite{Rusov2015,RusovArxiv2019} and Fig.~\ref{fig-lampochka}) and the strong toroidal field of the magnetic tube, $B_{tacho}^{Sun} = 4.1 \cdot 10^7 ~G$, as well as the average width of the ``thin'' ring $a_{conv} \sim 3.7 \cdot 10^{-4} ~H_p$ (see Eq.~(\ref{eq07-58}) and Fig.~\ref{fig-lower-heating}) between the O-loop and the walls of the magnetic tube (see Fig.~\ref{fig-lower-heating}), it is not difficult to show that the estimate of the Alfv\'{e}n speed (see Eq.~(\ref{eq07-63}))

\begin{equation}
v_A \cong 4.1 \cdot 10^6 ~cm/s
\label{eq07-64}
\end{equation}

\noindent allows us to estimate the analytical coincidence of the rise speed $v_{rise}$ (see Eq.~(\ref{eq07-61})) and the magnetic buoyancy speed $v_{B}$:

\begin{equation}
(v_{rise})_{conv} \equiv (v_{B})_{conv} \approx 1.4 \cdot 10^5 ~cm/s ,
\label{eq07-65}
\end{equation}

\noindent
at which for such large magnetic fields there is a significant number of rising tubes at all latitudes (see also the red lines in Figs.~\ref{fig-meridional-cut}a and~\ref{fig-magtube-tilt}a,b).

So, for the considered case, the secondary magnetic reconnection of the flux tubes leads to the real decrease in the magnetic field to $B \sim 10^5 ~G$ at $\sim 0.8 R_{Sun}$ (see Figs.~\ref{fig-magtube-tilt}a,b), and thereby reveals the nonzero magnetic buoyancy (see the blue lines in Fig.~\ref{fig-magtube-tilt}a,b). This means that for the condition of nonzero buoyancy ($\rho _{ext} \approx 0.09 ~g/cm^3$) and the toroidal magnetic field of the flux tube, as well as the transverse radius of the ``thin'' ring $\sigma _0 \equiv a_{conv} \sim 100 ~km \approx 4.5 \times 3.7 \cdot 10^{-4} H_p $ (see Fig.~\ref{fig-magtube-tilt}c,d), the estimate

\begin{equation}
v_A \cong 1.1 \cdot 10^4 ~cm/s
\label{eq07-66}
\end{equation}

\noindent
allows us to estimate the analytical coincidence of the rise speed $v_{rise}$ (see Eq.~(\ref{eq07-61})) and the magnetic buoyancy speed $v_{B}$:

\begin{equation}
(v_{rise})_{conv} \equiv (v_{B})_{conv} \approx 7.2 \cdot 10^2 ~cm/s ,
\label{eq07-67}
\end{equation}

\noindent
where we assume that MFTs are formed inside the disk at a height $z \sim H_p$ (see e.g. \cite{Khaibrakhmanov2017}). Below we consider the reason for the rising magnetic buoyancy 
only at low and middle latitudes.

%\begin{figure*}[tbp]

The second part of the scenario, which is based on the remarkable idea of \cite{Kichatinov1991} (see also Fig.~5 in \cite{Krivodubskij2005}), includes the generation of the magnetic field near the bottom of the convective zone (see Fig.~C.1b in \cite{RusovDarkUniverse2021} for the holographic BL~mechanism) and transfer of the toroidal field from the deep layers to the solar surface, where the efficiency of the magnetic buoyancy transfer is predetermined by the participation of two processes: macroscopic turbulent diamagnetism 
(see \cite{Zeldovich1957,Radler1968a,Radler1968b,Vainshtein1980,VainshteinKichatinov1983,Stix1989,Kichatinov1992,Kitchatinov2016}) and rotational $\nabla \rho$-pumping 
(see \cite{Drobyshevski1974,Vainshtein1983,VainshteinKichatinov1983,Stix1989,Kichatinov1991,Ossendrijver2002,Kitchatinov2012}), which are also associated with the process of meridional circulation 
(see e.g. \cite{Ballegooijen1982,Spruit1982,Dudorov1985,Ballegooijen1988,Wang1991,Choudhuri1995,Caligari1995,Nandy2002} and also \cite{Khaibrakhmanov2017}).

Let us note some important properties of macroscopic turbulent diamagnetism. It is known that \cite{Zeldovich1956,Zeldovich1957} and \cite{Spitzer1956} discovered the diamagnetism of inhomogeneously turbulent conducting liquids, in which the inhomogeneous magnetic field moves as a single whole. In this case the turbulent fluid, for example, with nonuniform effective diffusivity $\eta _T \approx (1/3) v l$ (see Fig.~1 in ~\cite{Kitchatinov2008}; see also Fig.~\ref{fig-Bz-diffusivity}b) behaves like a diamagnetic one and carries the magnetic field with the effective velocity

\begin{equation}
\vec{v}_{dia} = - \frac{1}{2} \nabla \eta_T ,
\label{eq07-68}
\end{equation}

\noindent
where $l$ is the mixing length of turbulent pulsations, and $v = \sqrt{\langle v^2 \rangle}$ is the root-mean-square velocity of turbulent motion. The minus sign on the right in Eq.~(\ref{eq07-68}) shows the meaning of turbulent magnetism: it is not paramagnetic magnetism, so magnetic fields repel from regions with relatively high turbulent intensity. In other words, macroscopic turbulent plasma diamagnetism and, as a consequence, the so-called macroscopic diamagnetic effect (see \cite{Radler1968a,Radler1968b}) in the physical sense is the displacement of the averaged magnetic field $B$ from regions with increased intensity of turbulent pulsations to regions with less developed turbulence \citep{Vainshtein1980,Krause1980}.

However, there is an interesting problem of the diamagnetic process caused by inhomogeneous turbulent intensity with allowance for the total nonlinearities in the magnetic field. This is due to the fact that up to the present time analytical estimates have been obtained only for the limiting cases of weak and strong magnetic fields. For example, at strong magnetic fields of flux tubes, at $B \gg B_{eq}$, the diamagnetic effect becomes almost negligible, in particular, strong magnetic damping of diamagnetism $\sim B^{-3}$ is obtained for super-equipartitions of fields \citep{Kichatinov1992}, when turbulence is close to two-dimensional \citep{Zeldovich1957}. On the other hand, for very weak fields the diamagnetic pumping, which is predetermined by the intensity of turbulence at $B \ll B_{eq}$, is a very effective process \citep{Kichatinov1992}.

Between the known limiting cases of weak and strong magnetic fields we are interested in the average toroidal magnetic field of a flux tube, that is $B _{tacho}^{Sun} \sim 10^5 ~G$, when $B \geqslant B_{eq}$. This is due to the fact that the secondary magnetic reconnection of the flux tubes (see Figs.~\ref{fig-upper-reconnection}b,d,f,g) leads to the real decrease of the magnetic field to $B \sim 10^5 ~G$ at $\sim 0.8~R_{Sun}$ (see the blue lines in Fig.~\ref{fig-magtube-tilt}a,b). This means that the real decrease in the toroidal magnetic field of the flux tube is a consequence of the formation of the
% reconnecting dynamo (see \cite{Baggaley2009}),
secondary reconnection,
as well as the important appearance of two ``anti-buoyancy'' effects: the downwardly directed turbulent diamagnetic transfer and the rotational effect of the magnetic $\nabla \rho$-pumping 
\citep{VainshteinKichatinov1983,Krivodubskij2005}.

In this regard, we consider turbulence with quasi-isotropic spectral tensor \citep{Kichatinov1987}, which is certainly the simplest representation for inhomogeneous turbulence (see Eq.~(2.12) in \cite{Kichatinov1992}. As a consequence, the information on the spectral properties of turbulence (given by Eqs.~(2.12)-(2.16) from \cite{Kichatinov1992}) is sufficient to reduce the expression for the average electromotive force $\varepsilon$ (see Eq.~(2.1) \cite{Kichatinov1992}) to its traditional form, where only integrations over the wave number $k$ and frequency $\omega$ remain. After such shortening it is possible to get (see \cite{Kichatinov1992})

\begin{equation}
\vec{F} = (\vec{v}_{dia} + \vec{v}_{dens}) \times \vec{B}
\label{eq07-69}
\end{equation}

\noindent
with the speed of turbulent diamagnetic transfer

\begin{equation}
\vec{v}_{dia} = -\nabla \int \limits _{0} ^{\infty} \Re _{dia} (k, \omega, B)
\frac{\eta k^2 q (k, \omega, x)}{\omega ^2 + \eta ^2 k^4} dk d\omega ,
\label{eq07-70}
\end{equation}

\noindent
where $q$ stands for the local velocity spectrum, and the rate of rotational magnetic advection caused by the vertical heterogeneity of the fluid density in the convective zone, i.e. the magnetic $\nabla \rho$-pumping effect,

\begin{equation}
\vec{v}_{dens} = \frac{\nabla \rho}{\rho} \int \limits _{0} ^{\infty} \Re _{dens} (k, \omega, B)
\frac{\eta k^2 q (k, \omega, x)}{\omega ^2 + \eta ^2 k^4} dk d\omega .
\label{eq07-71}
\end{equation}

\noindent
The effective speeds $v_{dens}$ and $v_{dia}$ are consequences of the non-uniformity of density and of turbulence intensity, respectively, where the latter is attributed to the known diamagnetic pumping.

We are interested in the problem of reconstructing a strong toroidal field $\sim 10^7 ~G$ of flux tubes (see Figs.~\ref{fig-meridional-cut}a and~\ref{fig-Bz-diffusivity}a), which by the secondary reconnection
 transform regions of the mean magnetic field $\sim 10^5 ~G$ in the convective zone (see Figs.~\ref{fig-meridional-cut}b and ~\ref{fig-Bz-diffusivity}a) and, thereby, allow the organization of the amazing balance between the magnetic buoyancy, turbulent diamagnetism, and the rotationally modified $\nabla \rho$-effect. It can be shown that for the parameters $\beta = B/B_{eq} \sim 1$ and $\cos \varphi$ (see Eq.~(3.5) in \cite{Kichatinov1992}), the speeds~(\ref{eq07-70}) and~(\ref{eq07-71}), which depend on the magnetic field $B$ through the kernels $\Re _{dens}(\beta, \varphi)$ and $\Re _{dia}(\beta, \varphi)$ (see Eqs.~(3.6) and~(3.7) in \cite{Kichatinov1992}), have the following estimates:

\begin{equation}
v_{dia} ^{black} \sim 1.3 \cdot 10 ~cm/s
\label{eq07-72}
\end{equation}

\noindent and

\begin{equation}
v_{dens} ^{light} \sim 7.2 \cdot 10^2 ~cm/s .
\label{eq07-73}
\end{equation}

This raises the question of how a certain balance appears between the speeds of magnetic buoyancy (see Eq.~(\ref{eq07-67}) and Fig.~\ref{fig-meridional-cut}), diamagnetic pumping (see Eq.~(\ref{eq07-72})), and rotating densely stratified $\nabla \rho$-pumping (see Eq.~(\ref{eq07-73}) and Fig.~\ref{fig-meridional-cut}b),

\begin{equation}
(v_B ^{red})_{conv} + v_{dia}^{black} + v_{dens}^{light} \cong 
(v_B ^{red})_{conv} + v_{dens}^{light} = ?
\label{eq07-74}
\end{equation}

In order to consider the balance Eq.~(\ref{eq07-74}), it is necessary to apply the widely used approximation of the mixing length (see e.g. \cite{BohmVitense1958,Bradshaw1974,Gough1977a,Gough1977b,Barker2014,Brandenburg2016}), which, according to \cite{Kichatinov1991}, fully satisfies this goal. This approximation will be understood as the replacement of nonlinear terms along with time derivatives in the equations for fluctuating fields by means of $\tau$-relaxation terms, i.e. instead of equations (3.1) and (3.9) from \cite{Kichatinov1991}, we now have the equation of the radial speed of the toroidal field in the convection zone

\begin{equation}
v_{dens}^{light} = \tau \langle u^2 \rangle ^{\circ}
\frac{\nabla \rho}{\rho} \left[ \phi_2 (\hat{\Omega}) - \cos ^2 \theta \cdot 
\phi_1 (\hat{\Omega}) \right] \approx
\label{eq07-75}
\end{equation}

\begin{equation}
\approx 6 v_p \left[ \phi_2 (\hat{\Omega}) - \cos ^2 \theta \cdot 
\phi_1 (\hat{\Omega}) \right] =
\label{eq07-76}
\end{equation}

\begin{equation}
= - \frac{3 \kappa g}{(\gamma - 1)c_p T}
\left[ \phi_2 (\hat{\Omega}) - \cos ^2 \theta \cdot \phi_1 (\hat{\Omega}) \right] ,
\label{eq07-77}
\end{equation}

\noindent
where $\tau \approx l / (\langle u^2 \rangle ^{\circ})^{1/2}$ is a typical lifetime of a convective eddy; $l$ is the mixing length; $\langle u^2 \rangle ^{\circ}$ is the mean intensity of fluctuating velocities for original turbulence; $\theta$ is the latitude; $e_r \nabla \rho / \rho = -e_r g / [(\gamma - 1) c_p T]$ , where $e_r$ is the radial unit vector; $T \cong 1.35 \cdot 10^6 ~K$ is the temperature at $0.8 R_{Sun}$; $g = g_0 (R_{Sun}/r)$ is the gravitational acceleration, where $g_0 = 2.74 \cdot 10^4 ~cm/s^2$ is the surface value; $c_p = 3.4 \cdot 10^8 ~cm^2 s^{-2} K^{-1}$ (fully ionized hydrogen) is the specific heat at constant pressure; $\gamma = 5/3$ is the ratio of specific heats $c_p / c_V$; $3 \kappa = \tau \langle u^2 \rangle ^{\circ}$, where $\kappa \equiv \eta _T \approx 5 \cdot 10^{13} ~cm^2 \cdot s^{-1}$ is turbulent diffusivity supplied by the model of non-rotating convection zone 
\citep{Spruit1974,Gough1976,Stix1990,Parker2009,Karak2014} (see also Fig.~1 in \cite{Kitchatinov2008}) and the mixing length relation $\langle u^2 \rangle ^{\circ} = - \nabla \Delta T l^2 g / (4 T)$, where the $\nabla \Delta T$ is superadiabatic temperature gradient; $v_p = (1/6) \tau \langle u^2 \rangle ^{\circ} (\nabla \rho / \rho)$ is the velocity of the magnetic field transfer caused by the density gradient (see Eq.~(36) in \cite{VainshteinKichatinov1983}); $\hat{\Omega} = Co = 2 \tau \Omega$ is the Coriolis number (reciprocal of the Rossby number), where $\Omega$ is the rotation speed, $\tau$ is the turnover time, and the functions

\begin{equation}
\phi _n (\hat{\Omega}) = (1/8) I_n (\Omega, k, \omega)
\label{eq07-78}
\end{equation}

\noindent
(see $I_1$ and $I_2$ in Eqs.~(3.12) and~(3.21) in \cite{Kichatinov1991}) are

\begin{equation}
\phi _1 (\hat{\Omega}) = \frac{1}{4 \hat{\Omega}^2} \left[
-3 + \frac{\hat{\Omega}^2 + 1}{\hat{\Omega}} \arctan \hat{\Omega} \right] ,
\label{eq07-79}
\end{equation}

\begin{equation}
\phi _2 (\hat{\Omega}) = \frac{1}{8 \hat{\Omega}^2} \left[
1 + \frac{\hat{\Omega}^2 - 1}{\hat{\Omega}} \arctan \hat{\Omega} \right]
\label{eq07-80}
\end{equation}

\noindent
(see also analogous Eq.~(19) and Fig.~2 in \cite{Kitchatinov2016}), which describe the rotational effect on turbulent convection.

According to \cite{Kapyla2014}, the rotational effect on the flow can be measured with the local Coriolis number $\hat{\Omega} = Co = 2 \tau \Omega$, especially if $\tau$ is estimated on the basis of the theory of mixing length, which predicts values of $\hat{\Omega}$ reaching more than 10 in the deep layers (see e.g. \cite{Ossendrijver2003,Brandenburg2005,Kapyla2011,KapylaEtAl2011}). However, on the other hand, according to \cite{Kapyla2014}, the question of whether there are such deep layers of solar and stellar convection zones or not is still open.

At the same time, we showed that this rotating magnetic $\nabla \rho$-pumping
appears precisely in the deep layers (through reconnection of the $\Omega$-loop
(see Figs.~\ref{fig-meridional-cut}b and~\ref{fig-Bz-diffusivity}) at middle and low latitudes, where the influence of the rotational gradient causes the upward magnetic pumping. The direction of the magnetic $\nabla \rho$-pumping (up or down, see Fig.~\ref{fig-meridional-cut}b) is susceptible to the sign of the factor in Eqs.~(\ref{eq07-79}) - (\ref{eq07-80}), which depends on the polar angle (colatitude) and the behavior of the Coriolis number functions in the convection zone.

Below we show that the estimate of the Coriolis number for solar convection in the deep layer should be $\hat{\Omega} \approx 20$. As a consequence of~(\ref{eq07-78})-(\ref{eq07-80}), the following values are assumed for this quantity:

\begin{equation}
\phi_1 \cong 0.0171 , ~~~ \phi_2 \cong 0.0098 ,
\label{eq07-81}
\end{equation}

\noindent
at which the radial velocity~(\ref{eq07-75}) of toroidal field transport changes the sign at the latitude $\theta ^{*} = \arccos \sqrt{\varphi_2 / \varphi_1} \cong 41^{\circ}$, being negative (downward) for $\theta > \theta ^{*}$ and positive (upward) for $\theta < \theta ^{*}$. Using~(\ref{eq07-77}), we find that the value of the radial velocity in the convective zone $v_{dens}^{light}$ (see Eqs.~(\ref{eq07-75})-(\ref{eq07-77})) near low latitudes (e.g. $\theta ^{*} = \arccos (0.985) \cong 10^{\circ}$;  see also Fig.~\ref{fig-meridional-cut}b) almost completely coincides with the value of the speed (\ref{eq07-73}), which was previously calculated with the help of the magnetic field $B$ in the kernel $\Re_{dens} (\beta, \varphi)$.

This means that the balance of the magnetic buoyancy (see Eq.~(\ref{eq07-67}) and Fig.~\ref{fig-meridional-cut}) and rotating density-stratified $\nabla \rho$-pumping (see Eq.~(\ref{eq07-73}) and Fig.~\ref{fig-meridional-cut}b) provides both the process of blocking the magnetic buoyancy at high latitudes,

\begin{equation}
\uparrow (v_B ^{red})_{conv} + \downarrow v_{dens}^{light} \leqslant 0
~~~ at~high~latitudes,
\label{eq07-82}
\end{equation}

\noindent
and the process of lifting MFTs from the base of the convective zone to the solar surface at low latitudes,

\begin{equation}
\uparrow (v_B ^{red})_{conv} + \uparrow v_{dens}^{light} > 0
~~~ at~lower~latitudes,
\label{eq07-83}
\end{equation}

\noindent
which are simultaneously predetermined by the following rise time value:

\begin{equation}
(\tau_B ^{red})_{conv} \sim \frac{z_0}{(v_B ^{red})_{conv} + v_{dens}^{light}}
\leqslant 4.5 ~~day,
\label{eq07-84}
\end{equation}

\noindent
where $z_0 \approx 0.1 ~H_p$ is the length of the magnetic line at $0.8 R_{Sun}$ (see the green band in Fig.~\ref{fig-Bz-diffusivity}). Here we must remember that magnetic $\nabla \rho$-pumping of plasma from the azimuthal field through the formation of the $\Omega$-loop is repeated again and again (see Fig.~\ref{fig-lower-reconnection} and Fig.~4 in \cite{Parker2009}). Moreover, our theoretical estimates of the rise time of the $\Omega$-loop are in good agreement with the experimental observations by \cite{Gaizauskas1983}, who mention the repeated appearance of new $\Omega$-loops with an interval of 5-8 days, which simultaneously indicates the continuing convective pumping of plasma.

We have shown above that the sunspots -- the dark magnetic regions occurring at low latitudes on the surface of the Sun (see Fig.~\ref{fig-meridional-cut}) -- are indicators of the magnetic field generated not with the help of the dynamo mechanism (this is very important!), which does not exist here, but with the help of the holographic BL~mechanism as components of the model of solar antidynamo (see Fig.~C.1b in \cite{RusovDarkUniverse2021}). Hence, the question arises as to how the sunspots originating from magnetic buoyancy on the surface only at low and, to a lesser extent, middle latitudes do not contradict the known observational data of the tilt angle of Joy's law (see Fig.~\ref{fig-magtube-tilt}c,d).

\begin{figure}
\begin{center}
\includegraphics[width=12cm]{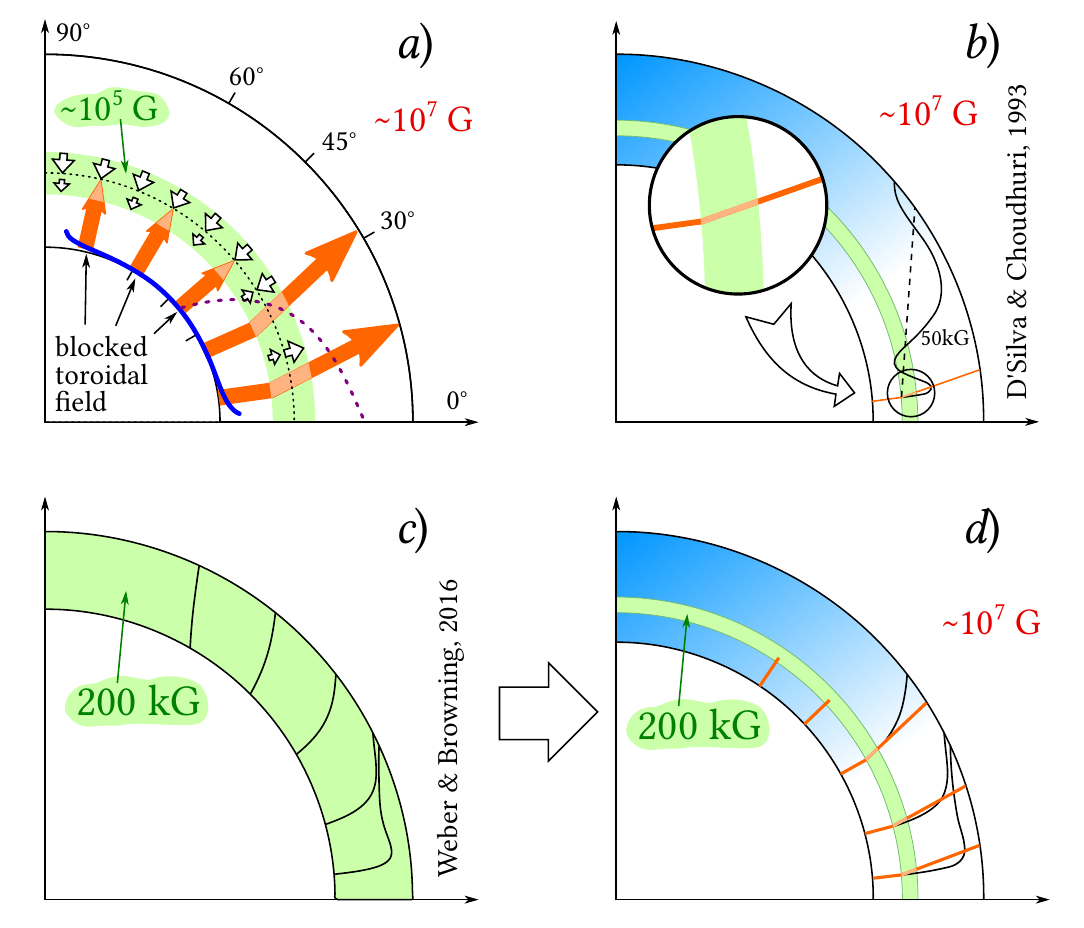}
\end{center}
\caption{\textbf{(a)} Interactions of magnetic buoyancy (red arrows) and rotating magnetic $\nabla \rho$-pumping (short white arrows) generate the total buoyancy of magnetic tubes in which the toroidal magnetic field $\sim 10^7 ~G$ predetermines the appearance of magnetic buoyancy by using the dominant Coriolis force in $\nabla \rho$-pumping with $\sim 10^5 ~G$ (see Fig.~\ref{fig-meridional-cut}b), which ultimately generates a curved upward loop on the surface of the Sun with a slope in the lower and, to a lesser extent, middle latitudes (see Fig.~\ref{fig-magtube-tilt}a,b). 
\textbf{(b)} According to \cite{DSilvaChoudhuri1993}, the trajectory of stream loops with 50~kG in the $\xi - \theta$ plane realized at latitude $5^{\circ}$ in the lower part of the convective zone. The dashed line shows the contour of constant angular momentum. A streaming ring of 50~kG, realized at $5^{\circ}$, also oscillates around this contour and comes out to a high latitude. The streaming ring ``hugs'' the contour of constant angular momentum, which is almost parallel to the axis of rotation. When we use Fig.~\ref{fig-meridional-cut-tilt}b, which is the modified Fig.~14 from \cite{DSilvaChoudhuri1993}, the inset shows the trajectory of the tubes with $\sim 10^7 ~G$, which, ultimately, by $\nabla \rho$-pumping with $\sim 5 \cdot 10^4 ~G$ (green line) generates a curved ascending loop on the solar surface with a tilt in the lower latitudes. At the same time, the flow ring ``hugs'' the contour of constant angular momentum, which is not parallel to the axis of rotation, which, of course, will be in good agreement with the observations of the slope angle of Joy's law (see Fig.~\ref{fig-magtube-tilt}c,d).
\textbf{(c)} According to \cite{Weber2016}, the trajectories of tubes initiated at $0.75 R_{Sun}$ (see Case T0 (black lines)) can cause (by means of induced Coriolis forces) tubes to move horizontally outwards to smaller layers at lower latitudes, and deeper layers at higher latitudes, caused by the direction of the flow backward.
\textbf{(d)} When we use Fig.~\ref{fig-meridional-cut-tilt}c, which is the modified Fig.~6 from \cite{Weber2016}, the trajectory of flux tubes with $\sim 10^7 ~G$ ultimately, by $\nabla \rho$-pumping with $\sim 5 \cdot 10^4 ~G$ (green line, see also Fig.~\ref{fig-meridional-cut-tilt}d) generates a curved ascending loop on the solar surface with a tilt in the lower latitudes.}
\label{fig-meridional-cut-tilt}
\end{figure}

Hence, it is very interesting that the holographic BL~mechanism, which is a satisfactory alternative theory against the action of the dynamo, predetermines not only the generation of sunspots themselves on the surface at low and middle latitudes (see Fig.~\ref{fig-meridional-cut}), but their coincidence with the observed slope angle of Joy's law, where, as a consequence, the average angle of inclination of bipolar sunspots increases with latitude (see Fig.~\ref{fig-magtube-tilt}c,d). This is due to the fact that strong toroidal magnetic fields in the overshoot tachocline are generated by the holographic BL~mechanism (see Fig.~C.1 in \cite{RusovDarkUniverse2021}), and bipolar magnetic tubes are created by lifting $\Omega$-loops caused by magnetic buoyancy. As a result, these ascending $\Omega$-loops, according to \cite{Choudhuri1989}, will be bent (towards the pole; see also Fig.~\ref{fig-meridional-cut-tilt}) by the Coriolis force, so that they eventually appear on the surface of the Sun with the tilt angle (see Fig.~\ref{fig-magtube-tilt}).

%\begin{figure}[tbp!]

Using the magnetic field strength near the bottom of the convective zone of the order of $10^5 ~G$ at $0.8 R_{Sun}$ (see Figs.~\ref{fig-meridional-cut}b and \ref{fig-Bz-diffusivity}a), we find that the Coriolis force plays a dominant role, and the MFTs, starting from the bottom at low latitudes, deviate by $0.8 R_{Sun}$ (see Figs.~\ref{fig-magtube-tilt}a,b and \ref{fig-meridional-cut-tilt}) and appear on the surface of the Sun at low and middle latitudes, located in the direction of the poles as sunspots. It is obvious that since the time of radiation diffusion of the flux tube (see Fig.~\ref{fig-magtube-tilt}a) is related to Eq.~(\ref{eq07-84}),

\begin{equation}
(\tau_B ^{red})_{conv} \equiv (\tau_d)_{conv} \leqslant 4.5 ~~day,
\label{eq07-85}
\end{equation}

\noindent
using the equations (\ref{eq07-63}), (\ref{eq07-66}), (\ref{eq07-67}), it is
possible to estimate the average width of the ``ring'' of the MFT:

\begin{equation}
a_{conv} \sim 100 ~km .
\label{eq07-87}
\end{equation}

Here we are interested in the relationship between the speed of magnetic buoyancy, $(v_B^{red})_{conv}$, and the average width of the ``ring'' $\sigma_0$, which is identical to $a_{conv}$ based on the vanBFF model (see Sect.~\ref{sec-radiative-heating}). Using Eq.~(23) from \cite{Choudhuri1987}

\begin{equation}
\frac{(v_B ^{red})_{conv}}{4.8 \cdot 10^3 ~cm/s} \equiv
u_t ' = \left( \frac{4 \pi}{C_D} \cdot \frac{\sigma_0}{R_{Sun}} \cdot 
\frac{(\Delta \rho / \rho_{ext})_0}{2 \cdot 10^{-6}} \right)^{1/2} ,
\label{eq07-88}
\end{equation}

\noindent
which connects the cross-section of the magnetic tube ``ring'' $\sigma_0 \equiv a_{conv}$ (see Eq.~(\ref{eq07-51})) with the thermal velocity in dimensionless coordinates (see analogous Table~1 in \cite{Choudhuri1987}), it is not difficult to show that for $(\Delta \rho / \rho_{ext})_0 \sim 10^{-5}$ (or equivalently at $B_0 = v_A(4 \pi \rho)^{1/2} \sim 10^5 ~G$) the magnetic buoyancy values $(v_B^{red})_{conv} / 4.8\cdot 10^3 ~cm/s \equiv u_t ' \approx 0.15$ and $\sigma _0 \equiv a_{conv} \sim 100 ~km$ remarkably coincide with the corresponding values of Eqs.~(\ref{eq07-67}) and~(\ref{eq07-87}), respectively.

\section{Summary and Outlook}
\label{sec-summary}

The main result of the theory of magnetic flux tubes is the properties of dark
matter axions identical to solar axions, the modulations of which are
controlled by the anticorrelated 11-year modulation of the asymmetric dark
matter (ADM) density in the solar interior~\citep{RusovDarkUniverse2021}:
\begin{enumerate}
\item
The existence of anchored flux tubes with $10^7 ~G$ in the overshoot tachocline is a consequence of the fundamental properties of the holographic principle of quantum gravity, one of which (unlike the dynamo action!) generates a strong toroidal field in the tachocline with the help of the holographic BL~mechanism (see Fig.~\ref{fig-solar-dynamos}b and C.1b in \cite{RusovDarkUniverse2021}).

\item
The theory of the almost empty anchored magnetic flux tubes with
$B \sim 10^7~G$ is the result of the formation of the magnetic O-loop (through
the primary reconnection) inside the MFT in the lower part of the convective
zone, which efficiently converts the solar axions and high-energy photons from
the radiation zone into photons of axion origin and axions of photonic origin,
respectively. The appearance of the axions of photonic origin is, on the one
hand, the manifestation of a ``ring'' of a strong magnetic tube due to
convective heating $(dQ/dt)_2$ (Fig.~\ref{fig-lower-heating}). On the other
hand, it leads to a remarkable result of the ``disappearance'' of Parker's convective
heat transport, and consequently, to the ris of the temperature in the lower part of the
magnetic tube. At the same time, the Parker-Biermann cooling effect is preserved in the
convective zone, when a free path opens for the photons of axion origin from
the tachocline to the photosphere! From here it is possible, for example, to obtain a
solution to the problem of corona heating~\citep{RusovDarkUniverse2021}.

The existence of magnetic buoyancy of flux tubes from the tachocline to the
surface of the Sun is a consequence of the primary trajectory formation through
the so-called universal model of van Ballegooijen-Fan-Fisher (vanBFF model),
which is determined by the equation $dQ/dt=(dQ/dt)_1+(dQ/dt)_2$. The first term
$(dQ/dt)_1$ defines the average temperature gradient between the lower
convective zone and the overshoot (see Eq.~\ref{eq07-57}), which deviates
significantly from the radiation equilibrium, which in turn suggests the
presence of a nonzero divergence of the heating radiation flux. The second term
$(dQ/dt)_2$ represents the convective diffuse radiation passing through a flux
tube due to temperature differences between the tube and the ambient plasma,
radiation thermal conductivity, and the mean width of the magnetic tube
``ring'' (see Eq.~(\mbox{\ref{eq07-58}}) and the area of the magnetic tube 
``ring'' $2\pi r a_{conv}$, where $r$ is the magnetic tube radius).

It was shown that at strong fields of $\sim 10^7 ~G$, which are determined by
the thermomagnetic EN effect (as a consequence of the holographic principle of
quantum gravity), the external gas pressure is almost equal to the magnetic
pressure between the tube and the ambient plasma. This leads to the virtually
empty magnetic tube: $dQ/dt \approx (dQ/dt)_2 \gg (dQ/dt)_1$. 
It means that solar axions, which are magnetically converted from the
high-energy photons of the radiation zone, on the one hand, directly suppress
the radiation heating in the convective zone, and on the other hand, initiate
an increase in convective heating by a sharp decrease in the average width of
the magnetic tube ``ring''. As a result, the convective heating $(dQ/dt)_2$
strongly dominates over the radiation heating $(dQ/dt)_1$.

This way we apply a new analysis of the universal vanBFF model, where the
calculated values such as the magnetic flux $\Phi$, the rise time $\tau_d$ and
the rise speed $(v_{rise})_{conv}$ of the MFT to the surface of the Sun, do not
contradict the known observational data.

\item
The existence of the magnetic buoyancy of flux tubes on the surface of the Sun is a consequence of the formation of a secondary trajectory by magnetic reconnection with $B \sim 10^5 ~G$ (see blue lines in Fig.~\ref{fig-magtube-tilt}a,b), at which, on the one hand, the formation of the rotating density-stratified $\nabla \rho$-pumping with $B \sim 10^5~G$ near the tachocline (see green bands in Figs.~\ref{fig-meridional-cut}b and ~\ref{fig-Bz-diffusivity}a) provides the process of blocking magnetic buoyancy at high latitudes, and on the other hand, it predetermines the existence of the dominant Coriolis force in $\nabla \rho$-pumping with $B \sim 10^5~G$, which ultimately generates an arched ascending loop on the surface of the Sun with a tilt in the lower and, to a lesser extent, middle latitudes (see Figs.~\ref{fig-magtube-tilt} and ~\ref{fig-meridional-cut-tilt}a).

\item
The basic properties of an almost empty magnetic tube (see Figs.~\ref{fig-lampochka} and ~\ref{fig-lower-heating}), characterizing thin radii of the cross-section ($\sigma_0 \equiv a_{conv} \sim 20~km$) and strong magnetic fields $\sim 10^7 ~G$, do not contradict the existence of a thin flux tube, since a virtually empty magnetic tube, unlike thin flux tubes, has not only strong magnetic fields, but also 
the width of the ``ring'' ($\sigma_0 \equiv a_{conv} \sim 100~km$) of the MFT
between the surface of the thick O-loop and the wall of the tube (see Figs.~\ref{fig-lampochka} and~\ref{fig-lower-heating}). It means that it is not the thin flux tubes with $\sim 10^5 ~G$ near the tachocline (see green bands in Figs.~\ref{fig-meridional-cut}b and~\ref{fig-Bz-diffusivity}a) that serve as a basis for the role of the Coriolis force at a certain latitude of the rising tilt in the direction of the active region. Therefore, the mean width of the magnetic tube ``ring''
(see Fig.~\ref{fig-lower-heating})
and the arched ascending loop of the magnetic field lines (see the tilt angle of Joy's law in Fig.~\ref{fig-magtube-tilt}c,d and Fig.~\ref{fig-meridional-cut-tilt}) are allowed.

\item
The averaged theoretical estimates of the magnetic cycle of flux tubes are practically identical to the observational (averaged) data of the tilt angle of Joy's law (see Fig.~\ref{fig-magtube-tilt}c,d).
\end{enumerate}

In the future, we will apply the results of the theory of magnetic tubes in strong fields not only to the Sun and a black hole, which was partially obtained in our work~\citep{RusovDarkUniverse2021}, but also consider its implications for the case of the well-known paradigm of fazzball complementarity in a black hole, which ``challenges'' the standard models of the central engine of a black hole for gamma-ray bursts and suggests a common physical mechanism behind GRBs, which indicates the magnetar central engine of GRBs.

%\section{Data availability}
%All the data used in this paper are publicly available, and all the
%corresponding references are given.

%%%%%%%%%%%%%%%%%%%% REFERENCES %%%%%%%%%%%%%%%%%%

\bibliographystyle{JHEP}

\bibliography{Rusov-AxionSunLuminosity}

\end{document}